\documentclass{aa} 

\usepackage{graphicx}
\usepackage{txfonts}
\usepackage[english]{babel}
\usepackage{amsmath}	% Advanced maths commands
\usepackage{amssymb}	% Extra maths symbols
\usepackage{multicol}        % Multi-column entries in tables
\usepackage{bm}		% Bold maths symbols, including upright Greek
\usepackage{pdflscape}	% Landscape pages
\usepackage{flafter}
\usepackage{wrapfig}
\usepackage[export]{adjustbox}
\usepackage{float}
\usepackage{enumitem}
\usepackage{tabularx} %for table size
\usepackage[normalem]{ulem}
\usepackage{subfig}
\usepackage{tablefootnote}

\usepackage[table,xcdraw]{xcolor}

\usepackage{afterpage}

\usepackage[colorlinks=true,linkcolor=blue,citecolor=blue,urlcolor=blue]{hyperref}

\begin{document} 
   
   \title{3D non-LTE modeling of the stellar center-to-limb variation for transmission spectroscopy studies}

   \subtitle{\ion{Na}{I} D and \ion{K}{I} resonance lines in the Sun}

   \author{G.~Canocchi\inst{\ref{inst1}} \and
          K.~Lind\inst{\ref{inst1}} \and 
          C. ~Lagae\inst{\ref{inst1}} \and
          A. G. M. ~Pietrow\inst{\ref{inst2}} \and
          A. M.~Amarsi\inst{\ref{inst3}} \and
          D. ~Kiselman\inst{\ref{inst4}} \and
          O. ~Andriienko\inst{\ref{inst4}} \and
          H. J.~Hoeijmakers\inst{\ref{inst5}}  
          %A. ~Brandeker\inst{\ref{inst1}}
          }
    \institute{Department of Astronomy, Stockholm University, AlbaNova University Center, SE-106 91 Stockholm, Sweden\\ \email{gloria.canocchi@astro.su.se}\label{inst1} \and 
    Leibniz-Institut für Astrophysik Potsdam (AIP), An der Sternwarte 16, 14482 Potsdam, Germany\label{inst2} \and
    Theoretical Astrophysics, Department of Physics and Astronomy, Uppsala University, Box 516, 751 20 Uppsala, Sweden\label{inst3} \and 
    Institute for Solar Physics, Dept. of Astronomy, Stockholm University, Albanova University Center, 106 91 Stockholm, Sweden\label{inst4} \and
    Lund Observatory, Department of Astronomy and Theoretical Physics, Lund University, Box 43, 221 00 Lund, Sweden\label{inst5} 
    }

% \abstract{}{}{}{}{} 
% 5 {} token are mandatory
 
  \abstract  
   {Transmission spectroscopy is one of the most powerful techniques to characterize transiting exoplanets since it allows to measure the abundance of the atomic and molecular species in the planetary atmosphere. However, the
   stellar lines can bias the determination of such abundances if their center-to-limb variations (CLVs) are not properly accounted for.}
  %3D NLTE analysis  
   {This paper aims to show that three-dimensional (3D) radiation hydrodynamic models and non-local thermodynamic equilibrium (non-LTE) line formation are required for an accurate modeling of the stellar CLV of the \ion{Na}{I} D$_1$ and \ion{K}{I} resonance lines on transmission spectra.}
   %Indeed it is well known that the commonly used 1D atmospheric models and LTE line formation fail to reproduce spatially resolved observations of the solar disk.}
  %methods
   {%We investigate the modeling of the CLV effect by comparing our synthetic spectra to high-spatial-resolution solar intensity atlases as well as to observations taken with the Swedish 1-m Solar Telescope.
   We model the CLV of the \ion{Na}{I} D$_1$ and \ion{K}{I} resonance lines in the Sun with 3D non-LTE radiative transfer. The synthetic spectra are compared to solar observations with high spatial and spectral resolution, including new data collected with the CRISP instrument at the Swedish 1-m Solar Telescope between $\mu=0.1$ and $\mu=1.0$.}
   {Our 3D non-LTE modeling of the \ion{Na}{I} D$_1$ resonance line at $5896\,\AA$ and the \ion{K}{I} $7699\,\AA$ resonance line in the Sun is in good agreement with the observed CLV in the solar spectrum. Moreover, the simulated CLV curve for a Jupiter-Sun system inferred with a 3D non-LTE analysis shows significant differences from that obtained from a 1D atmosphere. The latter indeed tends to overestimate the amplitude of the transmission curve by a factor that is of the same order of magnitude as a planetary absorption depth (up to 0.2\%).} 
   {This work highlights that in order to correctly characterize exoplanetary atmospheres, 3D non-LTE synthetic spectra should be used to estimate the stellar CLV effect in transmission spectra of solar-like planet hosts. Moreover, since different spectral lines show different CLV curves for the same geometry of the planet-star system, it is fundamental to model the CLV individually for each line of interest. The work will be extended to other lines and FGK-type stars, allowing synthetic high-resolution spectra to mitigate the stellar contamination of low-resolution planetary spectra, for example those from \textit{JWST}.}

   \keywords{Sun: atmosphere -- line: formation -- line: profiles -- Planets and satellites: atmospheres -- Techniques: spectroscopic 
            } 
\titlerunning{3D non-LTE modeling of the stellar CLV for transmission spectroscopy studies}
\authorrunning{Canocchi G. et al.}
   \maketitle

%-------------------------------------------------------%
%-------------------------------------------------------%
%------------------- INTRODUCTION ----------------------%
\section{Introduction}\label{sec: intro}

Transmission spectroscopy is a powerful tool to measure abundances of atomic and molecular species in the atmosphere of transiting exoplanets, especially close-in gas giants or Hot Jupiters (\citealt{Czesla2015}; \citealt{Casasayas2017}). When a planet transits across the stellar disk of its host star, the recorded stellar spectrum includes the planetary contribution. The latter can be detected as an extra absorption at specific wavelengths corresponding to the species that are present in the upper layers of the planetary atmosphere, through which the starlight is filtered during the transit. This technique works by subtracting the light recorded out of transit from that recorded during the transit, in order to detect the extra absorption due to the planet. In practice, high spectral resolution (i.e., $R \approx 100 \ 000$) %115000 for HARPS
is required to probe such weak features, which are usually of the order of only 0.01 - 0.1\%. %10$^{-3}$-10$^{-4}$. 
Therefore, high-resolution ultra-stable spectrographs such as  \textit{HARPS} (\citealt{Mayor2003}), \textit{HARPS-N} (\citealt{Cosentino2012}), \textit{CRIRES+} (\citealt{Kaeufl2004}; \citealt{Dorn2014}; \citealt{Dorn2023}), \textit{PEPSI} (\citealt{Strassmeir15}) or \textit{ESPRESSO} (\citealt{Pepe2021}) are needed for this kind of research.

To date, the most extensively studied atomic lines in transit spectroscopy are the lines of some alkali metals: the \ion{Na}{I} resonance doublet at 5890 $\AA\,$ and $5896\,\AA$, and the near-infrared \ion{K}{I} resonance line at 7699 $\AA\,$. High-resolution studies allowed to detect sodium in the atmospheres of several giant exoplanets, that are either Hot Jupiters or Neptunes, such as WASP-166b (e.g., \citealt{Seidel2020}), WASP-121b (e.g., \citealt{Hoeijmakers2020}; \citealt{Cabot2020};  \citealt{Seidel2023}), WASP-49b (e.g., \citealt{Wyttenbach2017}), WASP-69b (e.g., \citealt{Casasayas2017}), WASP-76b (e.g., \citealt{Seidel2019}), KELT-9b (e.g., \citealt{Hoeijmakers2019}), KELT-20b (e.g., \citealt{Casasayas2019}), HD 189733b (e.g., \citealt{Wyttenbach2015}; \citealt{Yan2017}) and HD 209458b (e.g., \citealt{Charbonneau2002}; \citealt{Snellen2008}). 
The latter is the first exoplanet where an atomic species was detected for the very first time by \citet{Charbonneau2002}. They detected neutral sodium through in-transit excess absorption in the \ion{Na}{I} D$_1$ and D$_2$ lines from observations with the Space Telescope Imaging Spectrograph (STIS) on board of the \textit{Hubble Space Telescope} (HST). After that, many other studies also from ground-based telescopes were carried out and several other species were detected including neutral potassium through the $7699\,\AA$ resonance line (\citealt{Casasayas2021}). The first detections of \ion{K}{I} occurred XO-2b (\citealt{Sing2011}) and in the eccentric hot Jupiter HD 80606b (\citealt{Colon2012}).

An important effect to consider when analyzing spectra with this technique is the center-to-limb variation (CLV) of the stellar line profiles. 
Indeed, \citet{Casasayas2020} stressed the importance of correctly accounting for this effect on the interpretation of the spectra of HD 209458.
The CLV describes the change in both strength and shape of the lines forming in the stellar atmosphere as the observer's line of sight moves from the center to the edge of the stellar disk.
 
By going towards the limb, the spectral lines are formed in higher layers of the photosphere. Different spectral lines are affected in different ways owing to their sensitivities on depth-dependent quantities such as temperature, electron pressure, and velocity, among others.  This CLV effect is closely related to continuum limb darkening: by going towards the limb, the higher, cooler layers of the photosphere are exposed to the observer, decreasing the observed surface brightness. 
The CLV of stellar lines can affect the correct determination of elemental abundances in the planetary atmospheres if not properly corrected for or even mimic planetary signals, thus leading to a false detection, since in some cases it is of the same order of magnitude as the planetary absorption feature (\citealt{Czesla2015}). During a transit, as the planet progresses, it obscures and blocks the light of parts of the stellar disk that contribute differently to the observed stellar line profile, which then varies among the different transit phases. Simply subtracting the out-of-transit light without considering the angle dependence will therefore introduce an error.

Although the CLV effect is recognized as an important and critical factor (e.g., \citealt{Muller2013}), it is often modeled by adopting the simple assumption of local thermodynamic equilibrium (LTE) and one-dimensional (1D) plane-parallel stellar atmospheres (e.g., \citealt{Casasayas2019}; \citealt{Chen2020}), or sometimes with 1D non-LTE models (e.g., \citealt{Yan2017}; \citealt{Casasayas2017}; \citealt{Morello2022}; \citealt{Sicilia2022}). 
However, several studies have shown that the commonly used 1D LTE models fail to reproduce spatially resolved observations of the solar disk. Specifically, the CLV of several atoms, such as Ca, C, O, Al and Fe lines, has been investigated in the Sun, finding that they are poorly modeled in 1D LTE, with errors on the line strength that range from strong (100\%) to moderate (10\%) at $\mu=0.2$ (e.g., \citealt{Pereira2009}; \citealt{Lind2017}; \citealt{Nordlander2017}; \citealt{Bjorgen2018}; \citealt{Amarsi2018b, Amarsi2019b}; \citealt{Bergemann2021}; \citealt{Pietrow2023paper, Pietrow2023letter}).

The \ion{Na}{I} doublet and the \ion{K}{I} 7699 $\AA\,$ resonance lines are strongly affected by departures from LTE; in particular over-recombination of the neutral species and photon losses in the resonance lines themselves (e.g., \citealt{Lind2011}; \citealt{Reggiani2019}). 
As such, 3D non-LTE modeling should be used for the highest accuracy.

The Sun is an excellent test bench for observing and studying the CLV of spectral lines (e.g., \citealt{Pierce1977}; \citealt{Neckel1994}).

Several solar atlases were recently produced and made publicly available, for example, the NSO/Kitt Peak FTS (\citealt{Stenflo2015}) and the Gregory Coud\'e Telescope SS3 (\citealt{Ramelli2017, Ramelli2019}) atlases, where the intensity of many spectral lines as a function of wavelength at different viewing angles ($\mu=\cos(\theta)$, with $\theta$ the heliocentric angle\footnote{$\theta$ is defined as the angle between the outward normal of the stellar atmosphere and the vector pointing from the center of the star toward the observer.}) from the limb (i.e. $\mu=0$) to the solar disk center (i.e. $\mu=1.0$) is recorded. Specifically, the FTS atlas extends in the range $4084-9950\,\AA$, but it samples only two positions along the solar disk: one at the center and one at the edge ($\mu=0.145$). Conversely, in the SS3 atlas the intensity at 10 different angles, spanning from $\mu=0.1$ to $\mu=1.0$ in steps of 0.1, is recorded but the covered spectral range is narrower ($4384-6610\,\AA$). Moreover, both these atlases show a noisy continuum (about 3\%) and are not corrected for telluric lines. 
Systematic and significant uncertainties in the continuum of different atlases of the solar disk-center intensity were also investigated in \citet{Doerr2016}.
An updated high-resolution atlas provided by the Institut f\"{u}r Astrophysik G\"{o}ttingen (IAG) that includes the CLV of several lines (in the wavelength range $4200-8000\,\AA$) has recently been published (\citealt{Ellwarth2023}). Nevertheless, it does not reach the very limb, $\mu=0.2$ being the farthermost angle sampled.

The CLV of stellar lines in stars other than the Sun has only been studied very recently. \citet{Dravins2017, Dravins2018} have tested how to resolve the surface of a K-type star, namely HD 189733A, using a transit of its exoplanet as a probe, and performing differential spectroscopy of a strong \ion{Fe}{I} line between different transit phases. This technique requires high-resolution spectra as well as very precise observations, but it remains a very promising candidate for spatially resolving the surface of many other stars in the near future. However, it should be noticed that the technique relies on the assumption that the planetary atmosphere contribution for that specific \ion{Fe}{I} line is null.

In this work, we study the 3D non-LTE CLV of absorption lines of neutral sodium and potassium in the Sun, %(Na) and (K)
and explore the implications for transmission spectroscopy studies.
In Sect.~\ref{sec: obs} we describe the observational data used in this work.
Complementing data from the literature,
we have carried out observations for the \ion{Na}{I} D$_1$ line at 5896 $\AA\,$ and the K\,I line at $7699\,\AA$ 
between $\mu=0.1$ and 1.0, 
obtained with the CRisp Imaging SpectroPolarimeter (CRISP; \citealt{Scharmer2008})
on the Swedish 1-m Solar Telescope (SST; \citealt{Scharmer2003}).
In Sect.~\ref{sec: methods} we describe our 3D non-LTE calculations, as well as our 3D LTE, 1D non-LTE, and 1D LTE calculations.
In Sect.~\ref{sec: analysis} we use these models to quantitatively analyze the solar observations, 
and demonstrate the impact of 3D non-LTE modeling on a simulated CLV transmission light curve in a Sun-Jupiter system.
Finally, in Sect.~\ref{sec: conclusions} we summarize the conclusions and discuss future perspectives.

%---------------------------------------------------%
%----------------- SOLAR OBSERVATIONS: ATLASES and SST -----------------%
\section{Observations and data processing}\label{sec: obs}
For this work, we employ data from the recently published solar intensity atlas obtained with the Fourier Transform Spectrograph (FTS; \citealt{Schafer2020}) at the IAG Vacuum Vertical Telescope (VVT), hereafter referred to as IAG FTS CLV Atlas, as well as observations from the SST/CRISP. The spectra from the IAG FTS CLV Atlas show a very high spectral resolution of about $R \approx 700 \ 000$ at $\lambda = 6000$ $\AA\,$, or $\mathrm{\Delta\nu=0.0024\,cm^{-1}}$ (\citealt{Reiners2023}) but are not corrected for telluric lines from water (H$\mathrm{_2}$O) and molecular oxygen (O$\mathrm{_2}$). Therefore, the regions affected by telluric lines were masked out. The IAG Atlas spans from $\mu=0.2$ to 1.0, in step of 0.1, with additional spectra at the following viewing angles: $\mu=0.35, 0.95, 0.97, 0.98, 0.99$, with a total of 14 $\mu$-angles covering the wavelength range $4200-8000 \ \AA\,$. These data are described in detail in \citet{Ellwarth2023}.

For comparison and completeness, we also took observations with the instrument CRISP at the SST of the \ion{Na}{I} D$_1$ resonance line at $5896 \ \AA\,$ and the \ion{K}{I} $7699 \ \AA\,$ resonance line. The datasets for these are described in Sect. \ref{sec: SST}, along with the reduction and calibration processes. A comparison between the two solar datasets, the IAG Atlas and the SST data, is presented in Sect. \ref{sec: comparison}.

\subsection{SST dataset}\label{sec: SST}
%%%%%%%%%%%%%%%%% TABLES %%%%%%%%%%%%%%%%%%
%%%%% Table 1: summary of the data %%%%% 
\begin{table*}%[ht]%[!b] or h!
\centering 
\caption{Summary of the SST/CRISP data. The columns show the date and time of observations, the campaign when they were taken, the observed line, the number of pointings in the scan (nP), the number of wavelength points (n$\lambda$), and the mean value of the solar elevation angle during observations.}\label{tab: SSTdata} 
\begin{tabular}{c c c c c c} \hline \hline 
Observation date, time (UT) & Campaign & Line & nP & n$\lambda$ & Elevation angle ($^{\circ}$)\\ \hline
2022-08-27, 08:15-10:41 & I & \ion{Na}{I} D$_1$ 5896 $\AA\,$ &  5 ($\mu=0.2, 0.4, 0.6, 0.8, 1.0$) & 39 & 36.8 \\ % Na I line: 22.3 (mu=0.2)-56.3 (mu=1)
2022-08-27, 08:15-10:41 & I & \ion{Fe}{I} 6173 $\AA\,$ &  5 ($\mu=0.2, 0.4, 0.6, 0.8, 1.0$) & 11 & 36.8\\ % Fe I line
2023-04-27, 17:23-18:30 & II & \ion{K}{I} 7699 $\AA\,$ & 3 ($\mu=0.2, 0.8, 0.9$) & 27 & 34.8\\ %K I line, day 1,34.8 at 18:00 from https://keisan.casio.com/exec/system/1224682277
2023-04-27, 17:23-18:30 & II & \ion{Fe}{I} 6173 $\AA\,$ & 3 ($\mu=0.2, 0.8, 0.9$) & 11 & 34.8 \\ %Fe I line, day 1
2023-04-28, 13:17-13:38 & II & \ion{K}{I} 7699 $\AA\,$ & 1 ($\mu=1.0$) & 27 & 72.8\\ %K I line, day 2
2023-04-28, 13:17-13:38 & II & \ion{Fe}{I} 6173 $\AA\,$ & 1 ($\mu=1.0$) & 11 & 72.8\\ %Fe I line, day 2
2023-04-29, 07:52-09:29 & II & \ion{K}{I} 7699 $\AA\,$ & 4 ($\mu=0.4, 0.6, 0.9, 1.0$) & 27 & 18.6\\ % K I line, day 3, AM=3.13
2023-04-29, 07:52-09:29 & II & \ion{Fe}{I} 6173 $\AA\,$ & 4 ($\mu=0.4, 0.6, 0.9, 1.0$) & 11 & 18.6\\ % Fe I line, day 3, taken at the same time so the airmass is the same
2023-04-30, 08:09-09:30 & II & \ion{K}{I} 7699 $\AA\,$ & 4 ($\mu=0.2, 0.4, 0.6, 0.8$) & 27 & 29.2\\ % K I line, day 4, 22.7-36.0, AM=2.1
2023-04-30, 08:09-09:30 & II & \ion{Fe}{I} 6173 $\AA\,$ & 4 ($\mu=0.2, 0.4, 0.6, 0.8$) & 11 & 29.2\\ % Fe I line, day 4
\hline 
\end{tabular}\\
\end{table*}
%%%%%%%%%%%%%%%%%%%%%%%%%%%%%%%%%%%%%%%%%%
%%%%%% Table with FPI tuning positions for each observed line %%%%%%
\begin{table}
\centering
\caption{FPI tuning positions for each observed line. The tuning positions are relative to a (daily) calibration on the line at disk center.} %added by Dan
\begin{tabularx}{\columnwidth}{X}
\hline
\hline
\ion{Na}{I} D$_1$ 5896 $\AA\,$ \\ \hline 
$-2000$, $-1740$, $-1480$, $-1220$, $-960$, $-700$, $-550$, $-440$, $-374$, $\pm$300, $\pm$248, $\pm$196, $-144$, $\pm$118, $\pm$92, $\pm$78, $\pm$66, $\pm$40, $\pm$14, $+140$, $+165$, $+381$, $+500$, $+578$, $+656$, $+812$, $+890$, $+968$, $+1070$ and 0 m$\AA\,$.\\ \hline
\hline
\ion{K}{I} 7699 $\AA\,$ \\ \hline 
$\pm$1380, $\pm$1150, $\pm$920, $\pm$690, $\pm$460, $\pm$230, $-178$, $\pm$138, $\pm$115, $\pm$92, $\pm$69, $\pm$46, $-23$, $+161$ and 0 m$\AA\,$.  \\ \hline
\hline
\ion{Fe}{I} 6173 $\AA\,$  \\ \hline 
$\pm$450, $\pm$175, $\pm$105, $\pm$70, $\pm$35 and  0 m$\AA\,$. \\ \hline
\end{tabularx}\label{tab:linesteps}
\end{table}
%%%%%%%%%%%%%%%%%%%%%%%%%%%%%%%%%%%%%%%%

%%%%%%%%%%%%%%%% FIGURES %%%%%%%%%%%%%%%%%%%%%
%%%% Figure with SST pointings %%%%
\begin{figure}
    \centering
    \includegraphics[width=8cm]{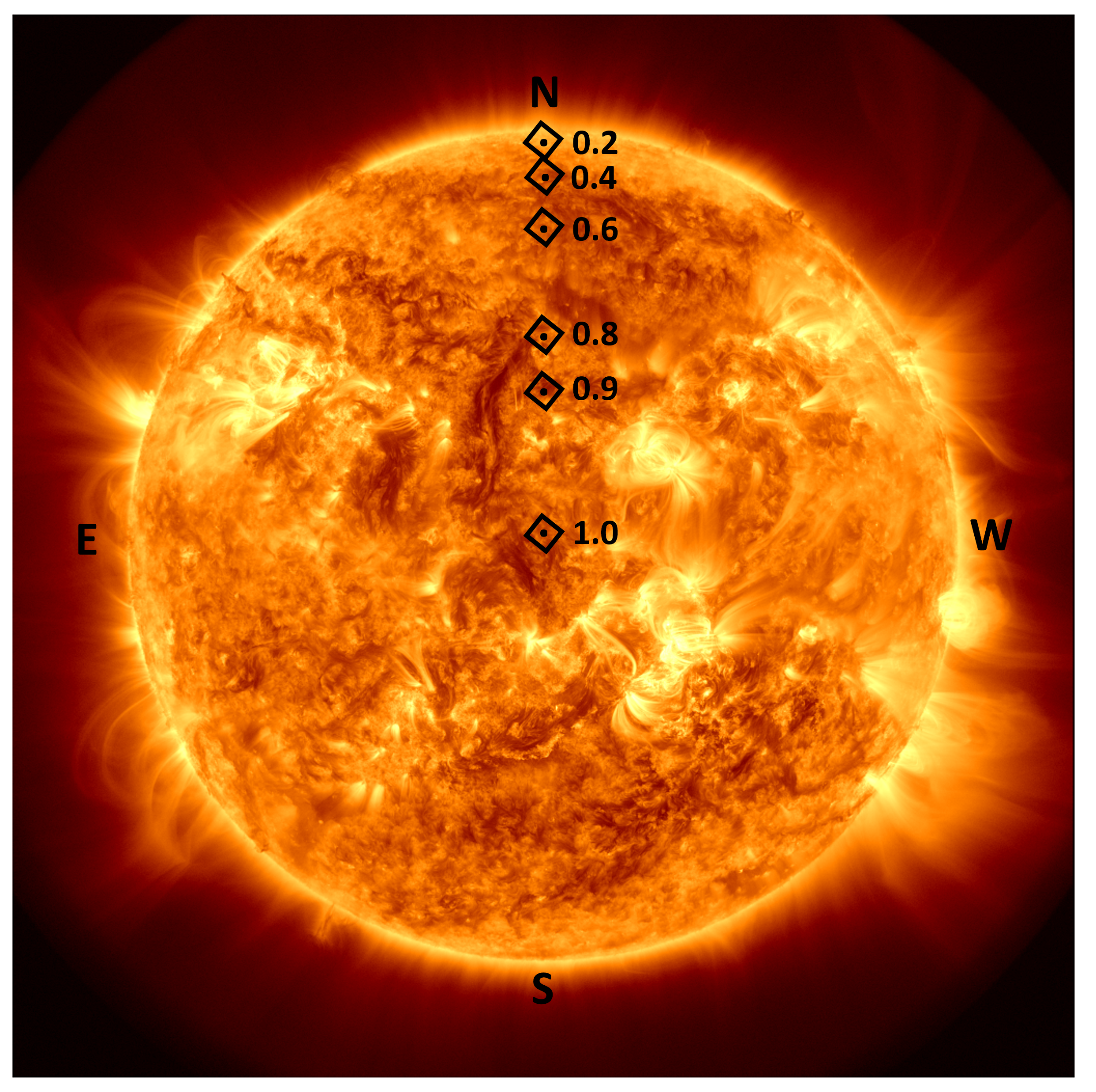}
    \caption{Full-disk image of the Sun on 29 April 2023 in the AIA 171 $\AA\,$ filter (\citealt{Lemen2012}). The black squares show all
telescope pointings used during the observation campaigns, spanning from the solar north pole to disk center.}  
    \label{fig: SSTpoint}
\end{figure}

We obtained the data for the \ion{Na}{I} D$_1$ and \ion{K}{I} lines in two campaigns, hereafter referred to as Campaign I and Campaign II, but following the same procedure. We pointed the telescope at several distinct positions on the disk, targeting quiet regions (i.e., without activity), from the solar north pole to the center, recording data for a time period of between 14 and 26 minutes at each position (i.e., $\mu$-angle). 
Since the tracking was active for the entire observation time, this method allows to perform temporal averages in order to correct for p-modes and oscillations. This is a well-established technique used in several previous studies (e.g., \citealt{Pierce1982}; \citealt{Ding1991}; \citealt{Pereira2009}; \citealt{Lind2017}). For the \ion{Na}{I} D$_1$ line, we pointed the telescope at $\mu=0.2, 0.4, 0.6, 0.8$ and 1.0, going along the meridian from the solar north pole to the disk center in order to avoid Doppler shifts from rotation. For the \ion{K}{I} line we also pointed at $\mu=0.9$, but the data from this pointing were not reduced in the end. 
The pointing positions are highlighted in Fig. \ref{fig: SSTpoint}. For $\mu=0.2, 0.4$ and 0.6, a wider $\mu$ range is covered by each pointing, due to the large Field of View (FoV) of the cameras. 
The FoV is circular, with a diameter of about $87 \arcsec$ on the solar disk. 

In addition to \ion{Na}{I} D$_1$ and \ion{K}{I} 7699 $\AA\,$ line, we also observed the photospheric magnetically sensitive line of \ion{Fe}{I} at 6173 $\AA\,$ in order to check the level of solar activity during measurements. This is the only line for which the polarimetric signal was measured.
A summary of the data is reported in Table \ref{tab: SSTdata} and the specific wavelength points sampled for each line are shown in Table \ref{tab:linesteps}. These positions are given relative to their line center and, especially for the \ion{Na}{I} D$_1$ line, they are not all symmetrically sampled\footnote{A value is symmetrically sampled if it has a $\pm$ symbol in front of it. It is not if it has only a $+$ or a $-$.}.

%First and second campaigns:
The observations were carried out under variable seeing conditions. This is acceptable because our first goal is to take large-scale averages along the FoV and use the fine-scale structure only for the alignment. Indeed, the only constraint was that there had to be at least one frame with high enough seeing to align it to co-temporal SDO/HMI observations, as explained hereunder. 
Campaign I was performed at the end of August 2022, and specifically, the \ion{Na}{I} D$_1$ and \ion{Fe}{I} 6173 $\AA\,$ data were taken between 08:15 and 10:41 UT on 27 August 2022 with SST/CRISP, as shown in Table \ref{tab: SSTdata}. We obtained one scan at 5 $\mu$-angles on the same day.
Just before the start of Campaign I, new detectors were installed at the SST, with a larger FoV compared to the previous set-up ($1.5\,\arcmin$ radius circular compared to the $1\arcmin \times 1\arcmin$ with the old detectors). The old detectors were from Sarnoff and the new ones are called "Ximea  CB262RG-GP-X8G3”. The new detectors also allowed to solve an issue with the back-scattering light in the red part of the spectrum. 
The data of Campaign I were the first to be taken with the new detectors.

Campaign II occurred at the end of April 2023, between April $27-30$ and we were not able to get one scan in one day due to a very variable seeing. Therefore, different pointings were performed each day depending on the seeing conditions (see Table \ref{tab: SSTdata}), but the data with the best seeing conditions were taken in the morning of 29 and 30 April 2023 and so these are the \ion{K}{I} data analyzed in this work. During this campaign we observed the \ion{K}{I} 7699 $\AA\,$ and \ion{Fe}{I} 6173 $\AA\,$ line, obtaining in total two scans at 6 $\mu$-angles (however, as previously mentioned, the data for $\mu=0.9$ are not used in this work). 

% Reduction process:
We reduced the data of both Campaigns with the SSTRED pipeline (\citealt{delaCruz2015}; \citealt{Lofdah2021}), which has been specifically designed to process the data from the SST. It corrects for dark and flat fields but also includes polarimetric calibrations and image restoration that removes optical aberrations due to turbulence in the atmosphere. This step is performed by the Multi-Object Multi-Frame Blind Deconvolution (MOMFBD; \citealt{Lofdah2002}; \citealt{VanNoort2005}) algorithm, which however can fail if the seeing conditions are poor (i.e., Fried parameter $r_0 \leq 5$ cm). This is not the case for our observations.

The SST pointing and tracking are good but in order to get a very precise estimate of the $\mu$-angle corresponding to each pixel of the FoV, we performed an additional aligning step. Several studies make use of cross-correlation algorithms to align their data (e.g., \citealt{reardon2012}) but they only work well when activity features like sunspots are clearly distinguishable in the FoV. For quiet Sun observations, this is not usually the case, especially at the limb with a lower image contrast. Therefore we use the same method as \citet{Pietrow2023paper}, which consists of manually aligning the images by looking for patterns among the granules and bright points that match with SDO/HMI continuum images (\citealt{Lemen2012}) which were enhanced in resolution and contrast by means of a deep learning algorithm (\citealt{DiazBaso2018}). Each HMI image was picked to be within half a minute of the highest seeing-quality scan within the first ten scans of each pointing. 
In this way, we can straightforwardly obtain absolute solar coordinates with respect to the disk center. After that, we assigned a $\mu$ value (see Eq. \ref{eq: mu}) at each pixel in the images by simply using the Pythagorean theorem to calculate the distance of each pixel from the disk center ($r$):
\begin{equation}
    \mu= \sqrt{1 - \left( \frac{r}{R_\odot} \right) ^{2} } 
    \label{eq: mu}
\end{equation}
where $R_\odot$ is the angular radius of the Sun as seen from the Earth in arcseconds. In Campaign I the solar radius is $R_\odot$= 949.5$\arcsec$ and it is 953$\arcsec$ in Campaign II\footnote{From \url{https://ssd.jpl.nasa.gov/horizons/app.html\#/}}.% The CRISP image scale is 0.044" pixels$^{-1}$.

Then, as previously mentioned, for the pointing at $\mu=$ 0.2, 0.4, and 0.6, a wider $\mu$ range is covered due to the large FoV of the cameras. Consequently, we were able to extract average line profiles for more $\mu$-angles. Specifically, from the pointing at $\mu=0.2$, we obtained the line profiles at $\mu=0.1, 0.15, 0.2,$ and 0.25. Similarly, from the pointing at $\mu=0.4$, we got the profiles at $\mu=0.3, 0.35, 0.4$ and 0.45. Lastly, from the pointing at $\mu=0.6$ we could also extract the profiles at $\mu=0.55$.
Finally, we obtain averaged line profiles for each spectral line, for the following 12 $\mu$-angles: 0.1, 0.15, 0.2, 0.25, 0.3, 0.35, 0.4, 0.45, 0.55, 0.6, 0.8, and 1.0. 
These are computed by averaging over time and over space all the pixels in the frames that are within the selected $\mu$ value $\pm$0.02. The CRISP image scale is 0.044$\arcsec$ pixels$^{-1}$.

% Intensity Calibration: 
After that, we performed another calibration of the disk center intensity (i.e., the $\mu=1.0$ line profile) using the ISPy\footnote{\url{https://github.com/ISP-SST/ISPy}} library (\citealt{ISPy2021}) by fitting several wavelength points of the averaged disk center profile to the IAG Atlas convolved with CRISP instrumental profile\footnote{The CRISP instrumental profile can be calculated with the "spec" module of ISPy.}. The calibration routine was modified because it usually employs the solar atlas by \citet{Neckel1984}. We then applied the intensity and wavelength calibration factors to all the other $\mu$-angle line profiles. This is again the same method employed in \citet{Pietrow2023paper}.

The resulting limb darkening curves and line profiles of the \ion{Na}{I} D$_1$ line taken in Campaign I and of the \ion{K}{I} line of Campaign II are shown in Appendix in Fig. \ref{fig: profilesNaD1} and \ref{fig: profilesK}, respectively. 

%----------------------------------------------------------%
\subsection{Comparison of solar data}\label{sec: comparison}
%%%%%%%%%%%%%% FIGURES %%%%%%%%%%%%%%%%%
%%%%%% Na D1 line profile at disk center (like Fig. 3 of APletter)%%%%%
\begin{figure} %From EW_computation.ipynb in SST_obs/
    \centering
    \includegraphics[width=9cm]{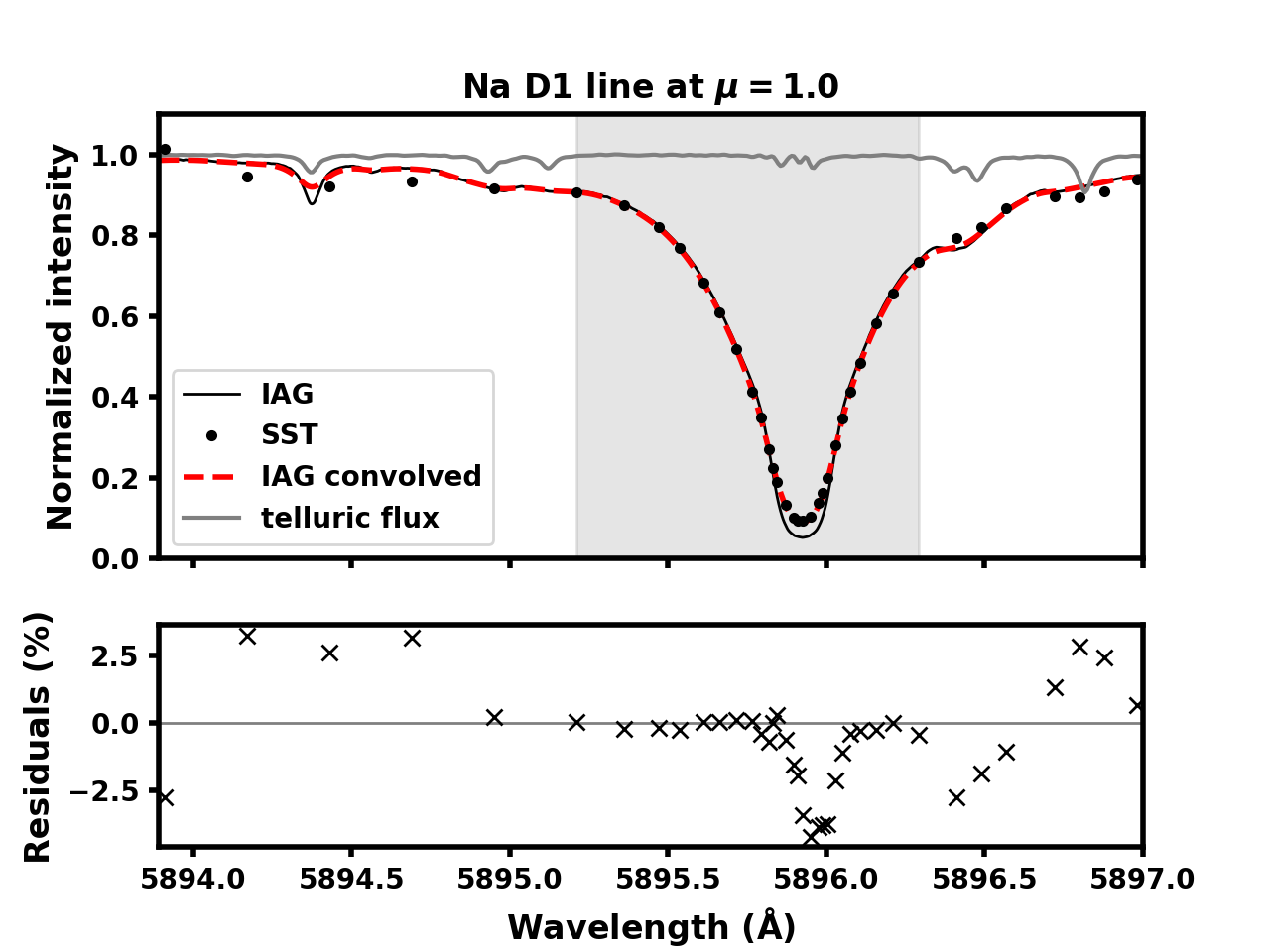} 
    \caption{\textbf{Top:} Comparison of the \ion{Na}{I} D$_1$ line profile at disk center between the SST/CRISP data (black dots) and the IAG Atlas convolved with the instrumental profile (red dashed line). The original profile from the IAG Atlas is also plotted with a black solid line. The grey-shaded region highlights the selected wavelength range for $W_\lambda$ computation. A telluric spectrum (\citet{Hinkle2000}) is overplotted as a grey line in the top panel, to show that there is negligible contamination in the $W_\lambda$ window. \textbf{Bottom:} the difference between the convolved IAG Atlas and the SST data in percentage.}
    \label{fig: SSTvsIAG_Na1}
\end{figure}
%%%%%%%%%%% K I line profile at disk center %%%%%%%%%%
\begin{figure}
    \centering
    \includegraphics[width=9cm]{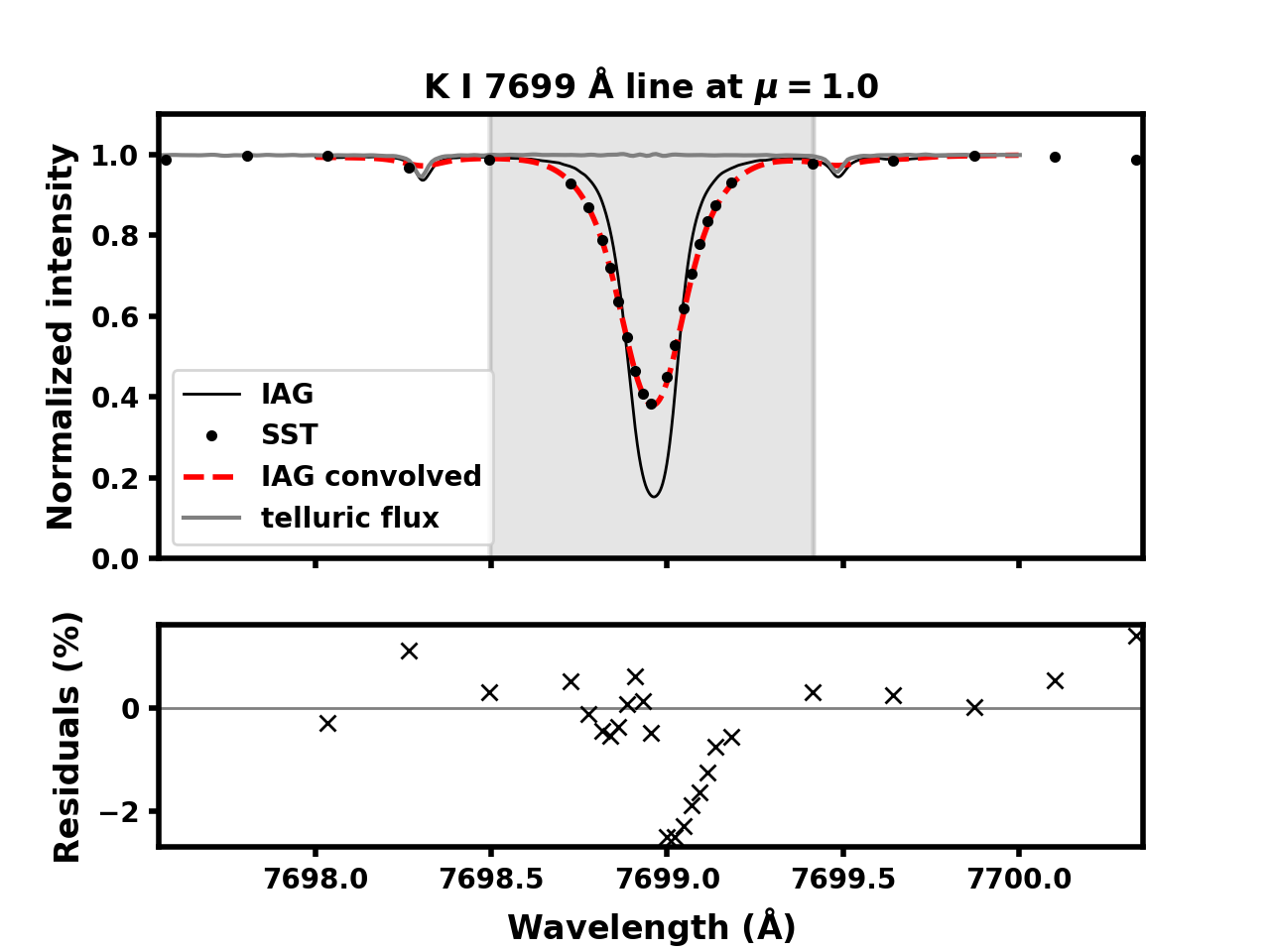}
    \caption{\textbf{Top:} Comparison of the \ion{K}{I} 7699 line profile at disk center between the SST/CRISP data (black dots) and the IAG Atlas convolved with the instrumental profile (red dashed line). The original profile from the IAG Atlas is also plotted with a black solid line. The grey-shaded region highlights the selected wavelength range for $W_\lambda$ computation. A telluric spectrum (\citet{Hinkle2000}) is overplotted as a grey line in the top panel, to show that there is negligible contamination in the $W_\lambda$ window. \textbf{Bottom:} the difference between the convolved IAG Atlas and the SST data in percentage.} 
    \label{fig: SSTvsIAG_K1}
\end{figure}
%%%% Figure: EWs of Na D1 comparison between SST and IAG atlas %%%% max percentage diff is 2.8 % at mu=0.2
\begin{figure}
    \centering
    \includegraphics[width=10cm]{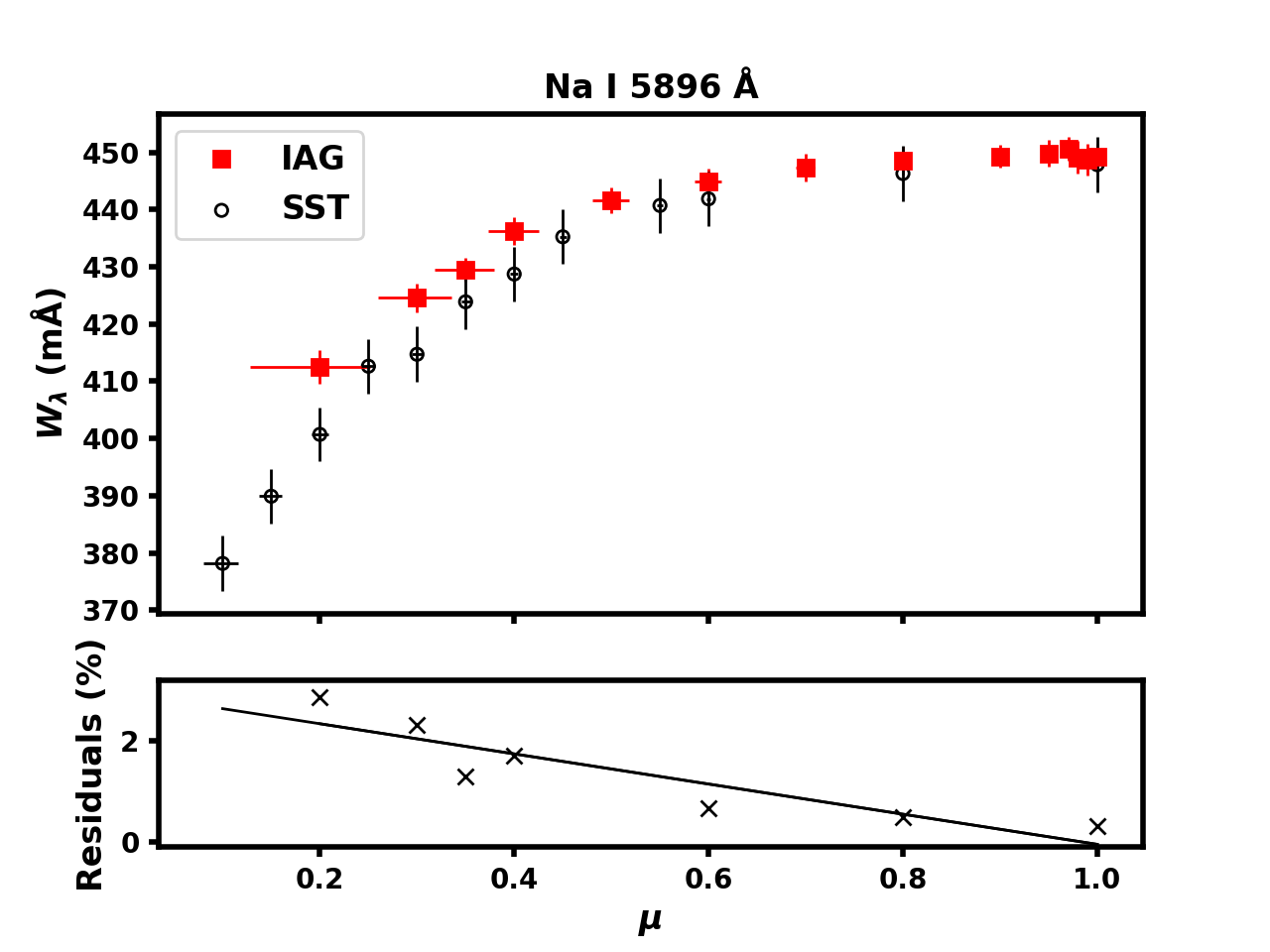} 
    \caption{\textbf{Top:} Equivalent width of the \ion{Na}{I} D$_1$ line as a function of the $\mu$-angle. Comparison of the SST/CRISP data (black dots) with the IAG FTS CLV Atlas (red squares) convolved with CRISP instrumental profile. \textbf{Bottom:} the difference between the IAG and the SST data in percentage.}  
    \label{fig: SSTvsIAG_Na2}
\end{figure}
%%%%%%%%%%%%%%%%%%%%%%%%%%%%%%%%%%%%%%%%%%%%%%%%%%%
%%%% Figure: K I comparison between SST and IAG atlas %%%% max percentage difference is 1.9% at mu=0.8
\begin{figure}
    \centering
    \includegraphics[width=10cm]{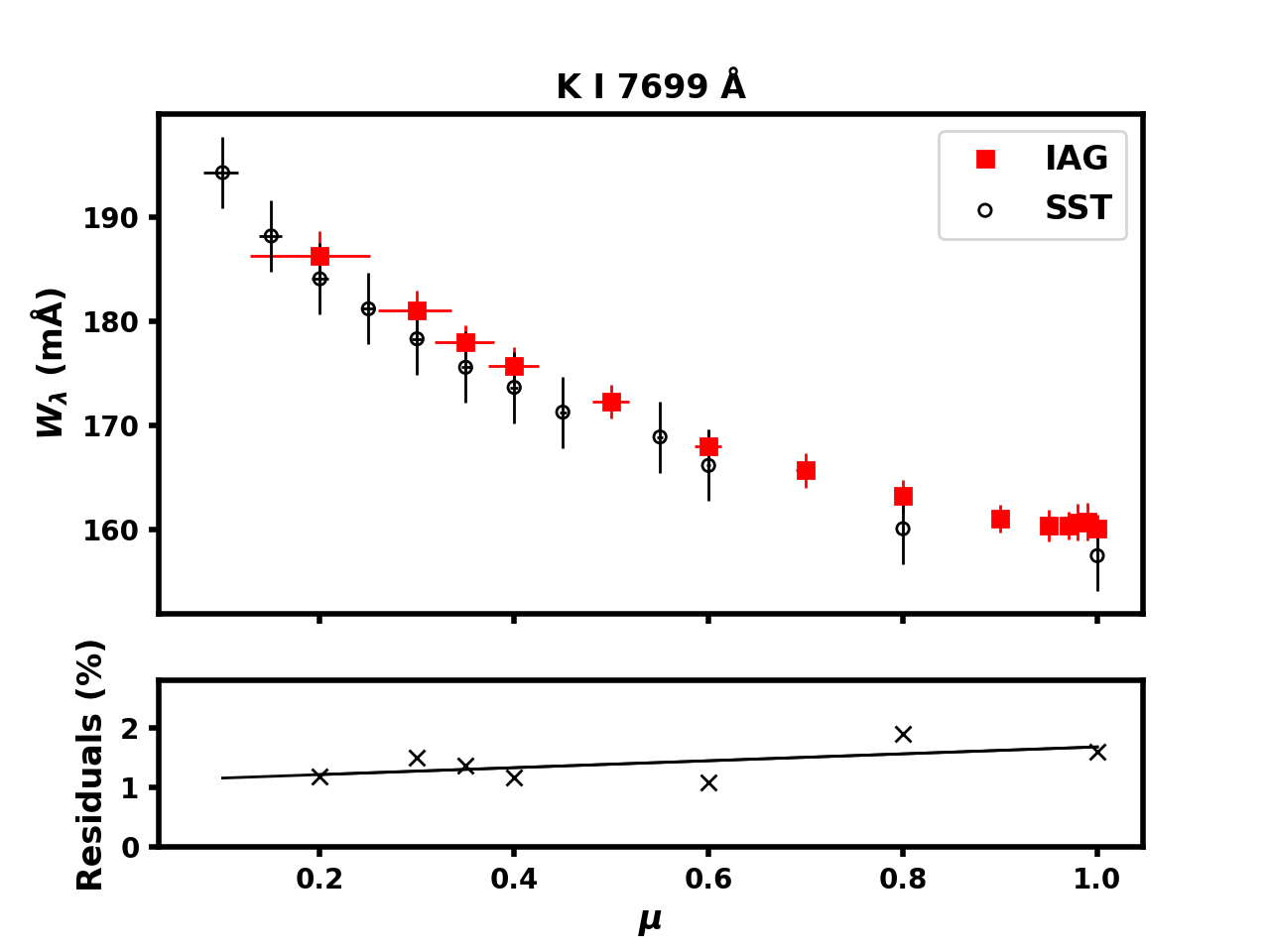}
    \caption{\textbf{Top:} Equivalent width of the \ion{K}{I} line at 7699 $\AA\,$ as a function of the $\mu$-angle. Comparison of the SST/CRISP data (black dots) with the IAG FTS CLV Atlas (red squares) convolved with CRISP instrumental profile. \textbf{Bottom:} the difference between the IAG and the SST data in percentage.}
    % from EWs_K_lines2.ipynb
    \label{fig: SSTvsIAG_K}
\end{figure}
%%%%%%%%%%%%%%%%%%%%%%%%%%%%%%%%%%%%%%%%%%%%%%%%%%%
%--------------------------------------% 
In this section, we compare our new SST/CRISP observations to the IAG FTS Atlas through the measurement of equivalent widths ($W_\lambda$) at different $\mu$-angles.
First of all, we convolved the Atlas with CRISP instrumental profile in order to degrade it to the same resolution as our SST data, which is about $R \approx 160 \ 000$ at $7699\,\AA$. 
Such convolution is shown in Fig. \ref{fig: SSTvsIAG_Na1} and \ref{fig: SSTvsIAG_K1} where the line profile of the \ion{Na}{I} D$_1$ and \ion{K}{I} line, respectively, at disk center (i.e., $\mu=1.0$) is compared between the two datasets. From these figures, it is apparent that the agreement between the two datasets at disk center is excellent, and specifically, it is within 2.5\% for the \ion{K}{I} line and within 3\% for the \ion{Na}{I} D$_1$ line.

Next, we computed the $W_\lambda$ of all the Na line profiles at different $\mu$-angles in the wavelength range 5895.21 -- $5896.291\,\AA\,$ (shaded region in Fig. \ref{fig: SSTvsIAG_Na1}) in order to exclude the blending on the right wing due to an uncorrected telluric line that is clearly visible in Fig. \ref{fig: SSTvsIAG_Na1}. It is worth noting that we are not integrating the entire line profile of the \ion{Na}{I} D$_1$ line up to the continuum. 
The $W_\lambda$ of the \ion{Na}{I} D$_1$ line as a function of the $\mu$-angle is shown in Fig. \ref{fig: SSTvsIAG_Na2}.

% Rephrasing of the previous sentence:
From the latter, it is apparent that the $W_\lambda$ of the SST data is slightly lower than the $W_\lambda$ of the corresponding IAG profile. Even with this small difference linearly increasing towards the limb, the agreement between the two datasets is quite good, and specifically, it is within  0.3\% at disk center ($\mu=1.0$) and about 2.9\% at the limb ($\mu=0.2$). It is worth noting that the SST pointings are more precise than the IAG, especially at the limb because of the method we used to obtain the $\mu$-position of each pixel of the frames, that is the alignment with simultaneous space-based data (i.e., SDO/HMI images) which translates into sub-arcseconds precision at the limb. This procedure is accurately described in Sect. \ref{sec: SST}.  Conversely, the limitations of the IAG Atlas come from the use of fibers, which result in increasingly large uncertainties in the $\mu$-positions closer to the limb, as is apparent from the horizontal error bars in Fig. \ref{fig: SSTvsIAG_Na2}.
%%%%%%%%%%%%%%%%%%%%%%%%%%%%%%%%%%%%%%%

The same calculations were performed for the \ion{K}{I} line, computing the $W_\lambda$ in the wavelength range 7698.495 -- $7699.415\,\AA\,$, as highlighted in Fig. \ref{fig: SSTvsIAG_K1}. The resulting CLV of this line is shown in Fig. \ref{fig: SSTvsIAG_K}, where $W_\lambda$ at different $\mu$-angles are compared between the IAG and SST data. Also in this case the $W_\lambda$ of the SST data is systematically lower than the IAG data but overall they agree within $1-2$\% at all $\mu$-angles. Differences in continuum fitting may have contributed to this difference in line strength, but we emphasize that the atlases everywhere agree within the estimated pointing errors.  

%%%% EWs uncertainties %%%%
With regards to the uncertainties on the IAG datapoints, the horizontal error bars on $\mu$-angles were provided by Ellwarth (priv. comm.) whereas the vertical error bars on equivalent widths were estimated from Cayrel's formula (\citealt{Cayrel1988}; \citealt{Cayrel2004}). They are reported in Table \ref{tab: EWIAG} in Appendix. 
The pointing of the SST is very precise; we estimate the horizontal error bar on $\mu$ from the aligning procedure of SST frames with SDO images described in Sect. \ref{sec: SST}. Indeed, the geocentric orbit of SDO leads to a potential error of maximum $1.7 \arcsec$. In $\mu$-space, this translates into a negligible value at disk center (about $10^{-6}$) but becomes significant at the limb, reaching 0.02 at $\mu=0.1$.
The vertical error bars on $W_\lambda$ of the SST data correspond to the standard deviation of the individual pointings ($\sigma_\mu$) added to an estimated error due to continuum placement ($\sigma_\mathrm{continuum}$), which represents the main source of uncertainty for this dataset. These errors were computed by vertically shifting the continuum within 1\% and estimating how the $W_\lambda$ change. Therefore the total error becomes:
\begin{equation}
    \sigma^2_\mathrm{tot}= \sigma^2_\mu + \sigma^2_\mathrm{continuum}
    \label{eq: sigmaSST}
\end{equation}
These values are shown in Table \ref{tab: EWSST} in Appendix. 

%--------- SYNTHETIC SPECTRA ---------------%
\section{Radiative transfer calculations}\label{sec: methods}
\subsection{Synthetic spectra}\label{sec: model}
%%%%%%%%%%%%%% FIGURES %%%%%%%%%%%%
%%% EW NLTE/LTE %%%
 \begin{figure}
 \centering
   \includegraphics[width=9cm]{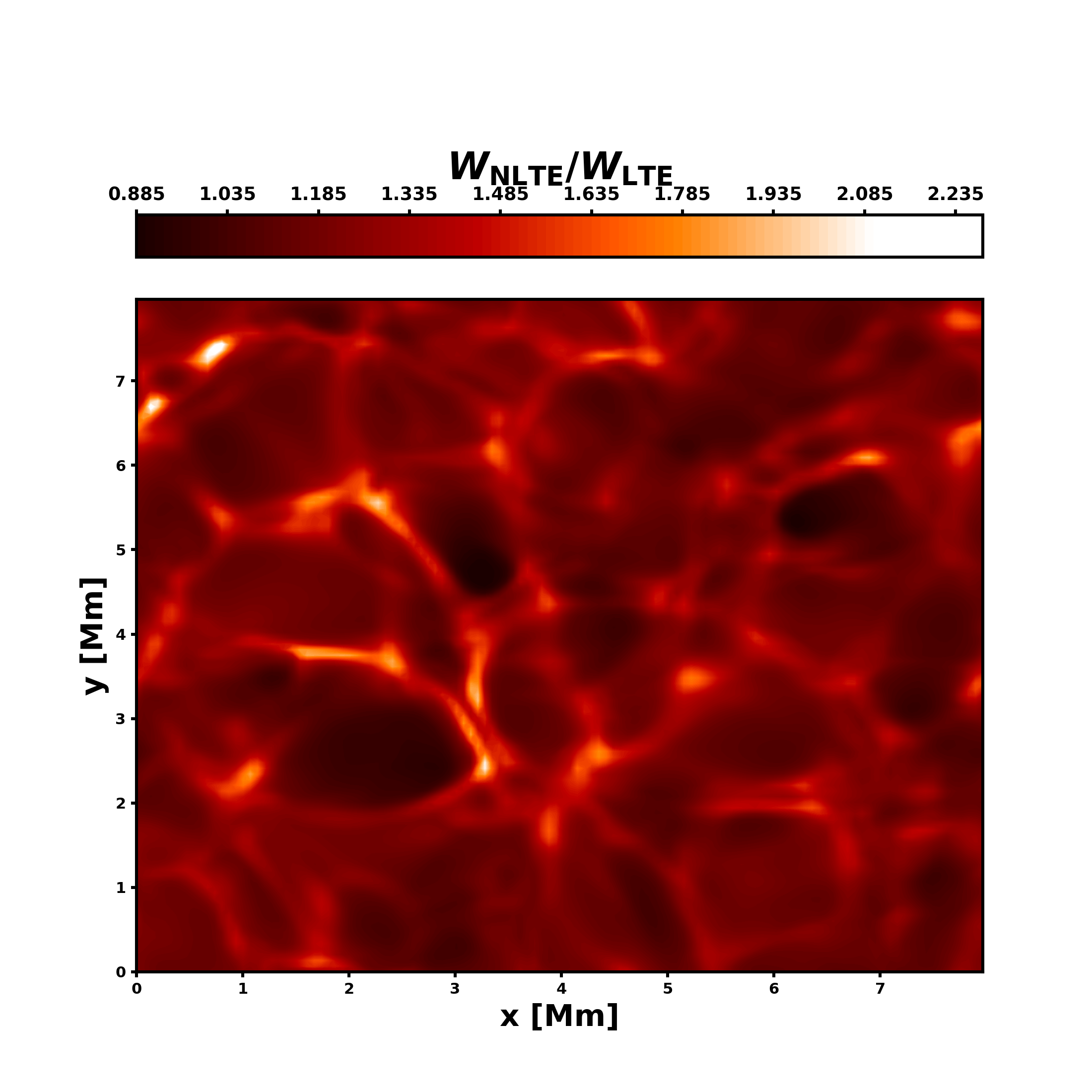}
      \caption{Ratio of the NLTE to LTE equivalent width of \ion{K}{I} $7699 \ \AA\,$ for a single \texttt{Stagger} snapshot. In the up-flowing granules, overionization makes the line weaker in NLTE, whereas in the intergranular lanes the NLTE $W_\lambda$ is stronger.}
         \label{fig: stagger}
   \end{figure} 
%%%%%%%%%%%%%%%%%%%%%%%
%%%%%%%%%% TABLES %%%%%%%%%%%%
\begin{table*}[h]
\centering  
\caption{Atomic data of the sodium and potassium lines considered in the CLV analysis, as implemented in the model atoms of \citet{Lind2011} and \citet{Reggiani2019}. The last column shows the wavelength windows used in the integration for the equivalent width computation.}\label{tab: atomicdata} 
\begin{tabular}{c c c c c c c} \hline \hline 
$\lambda$ (\AA) & Transition & $^{(a)}\Delta$E (cm$^{-1}$) & $^{(b)}$log ($gf$) & $^{(c)}\sigma$ (a.u.) & $^{(d)}\alpha$ & $\lambda_\mathrm{int}$ (\AA)\\ \hline
5896 & $\mathrm{3s  ^2S_{1/2}-3p^2P^0_{1/2}}$ & 16956.17 & -0.194 &  407 & 0.237 & 5895.2100-5896.2910\\ %Na D1
5688 & $\mathrm{3p  ^2P^0_{3/2}-4d^2D_{5/2}}$ & 17575.36 & -0.452 & 1955 & 0.327 &  5688.0546-5688.3546\\ % VALD and model atom do not agree
6154 & $\mathrm{3p  ^2P^0_{1/2}-5s^2S_{1/2}}$ & 16244.50 & -1.547 & - & - & 6154.1053-6154.3453\\
6160 & $\mathrm{3p  ^2P^0_{3/2}-5s^2S_{1/2}}$ & 16227.30 & -1.246 & - & -  & 6160.6270-6160.8670\\
7699 & $\mathrm{4s^2S_{1/2}-4p^2P^0_{1/2}}$ & 12985.18 & -0.176 & 486 & 0.232 & 7698.4953-7699.4153\\ \hline %K I
\end{tabular}\\
\begin{flushleft}
\textbf{Notes.}\\
$^{(a)} \Delta E$ is the energy difference between the levels of the transition, from the NIST database\tablefootnote[6]{\url{https://physics.nist.gov/PhysRefData/ASD/lines_form.html}} (\citealt{nist}).\\
$^{(b)}$ log ($gf$) are from the VALD database\tablefootnote[7]{\url{http://vald.astro.uu.se/}} (\citealt{vald}), where $f$ is the oscillator strength, or transition probability and $g$ is the statistical weight.\\
$^{(c)} \sigma$ is the broadening cross-section for elastic collisions with hydrogen at a relative velocity of 10$^4 \mathrm{m \ s^{-1}}$.\\
$^{(d)} \alpha$ is the exponent with which the cross-section $\sigma$ varies with velocity, that is $v^{-\alpha}$ (\citealt{Anstee1995}).\\
\end{flushleft}
\end{table*} 
%------------------------------------------------------------%

The synthesis of the solar spectrum was performed with the radiative transfer (RT) MPI-parallelised code \texttt{Balder} (\citealt{Amarsi2018}), an extensively modified version of the \texttt{Multi3d} code (\citealt{Botnen1999}; \citealt{Leenarts2009}), which can solve the restricted non-LTE problem for trace elements in 1D and 3D, for user-specified atomic elements. 
\texttt{Balder} employs the \texttt{BLUE} (\citealt{Amarsi2016, Amarsi2016b}) opacity package, which computes the line opacity on a grid of wavelength, density and temperature. A full 3D non-LTE analysis of several elements in the Sun, including Na and K, was performed with the same code by \citet{Asplund2021}.

For the model atmosphere, a 3D radiative-hydrodynamics simulation with 
the \texttt{Stagger}-code (\citealt{Galsgaard1995}; \citealt{Stein1998}; 
\citealt{Collet2011}) was used.
\texttt{Stagger} employs the "box-in-a-star" set-up, meaning that it simulates a rectangular box, vertically open, with periodic horizontal boundaries, located in the upper part of the solar photosphere, without including any chromospheric layer.  
In this work, we simulate the quiet Sun, without including magnetic fields\footnote[8]{\texttt{Stagger} can compute M-RHD including magnetic fields, but we do not use this feature here.}.
The 3D model is set on a cartesian grid of size 240$\times$240$\times$240, corresponding to a physical size of 8$\times$8$\times$4 Mm. The vertical dimension is characterized by non-equidistant spacing in order to obtain a higher resolution at the regions of interest and to resolve the strong temperature gradient of the photosphere. In order to decrease the computational time for the spectral synthesis, all the atmospheric snapshots were resized to a resolution of 60$\times$60$\times$101, from their original value of 240$^3$. This approach has been already tested in previous studies, such as \citet{Lind2017} and \citet{Lagae2023}.

Ten snapshots were used, from the same simulations discussed already in several earlier analyses of the Sun
(\citealt{Amarsi2018b,Amarsi2019b,Amarsi2020,Amarsi2021};
\citealt{Asplund2021}).  We note that this simulation is more recent than that in the \texttt{Stagger}-grid (\citealt{Magic2013}), with improvements to the opacity binning scheme (\citealt{Collet2018}) and extending for 
over one day of solar time. 
The mean effective temperature of these snapshots is $T_\mathrm{eff}=5770 \pm 5\,K$\footnote[9]{The adopted value is close to the nominal $T_\mathrm{eff}= 5772\rm\,K$ (\citealt{Prsa2016})}.

An example of the surface image of one of the snapshots of the solar photosphere selected for this work is shown in Fig. \ref{fig: stagger}. It represents the 3D non-LTE to 3D LTE ratio of $W_\lambda$ of the \ion{K}{I} $7699\,\AA$ line, showing that the non-LTE effects vary with the convection pattern: under-population in the up-flowing granules and the opposite effect, over-population, in the intergranular lanes. Such a variation is expected and is due to the much steeper temperature gradients of the granules compared to those of the intergranular lanes. The granulation pattern of the solar photosphere, which arises from the convective motion of the hot plasma and gas, is clearly recognizable in this figure. This is a natural 3D effect that leads to spectral line broadening and line asymmetries (e.g., \citealt{Dravins1981}). 
In the 3D non-LTE models the broadening effects of these temperature inhomogeneities and convective velocity structures are naturally included, therefore no extra broadening is required.  In 1D, however,
the line broadening is instead roughly accounted for by the tunable
parameters microturbulence ($v_\mathrm{mic}$) and macroturbulence. For the Sun, in the literature a microturbulence of 1.0\,km s$^{-1}$ is usually adopted, and this was the one employed in this work for the 1D simulations, independent of depth or viewing angle. For the macroturbulence velocity ($v_\mathrm{mac}$) a value of 3.5\,km s$^{-1}$ was used. The 1D synthetic spectra have been computed with the \texttt{MARCS} (\citealt{Gustafsson2008}) solar atmosphere. 

%%%%%% Line profile analysis %%%%%% 

The synthetic spectra were computed at three different abundances for each snapshot. Then the line profile was spatially and temporally averaged, using all ten snapshots. After that, the disk-center line profile was fitted to the observed data via a standard $\chi^{(2)}$ minimization routine. Linear interpolation between models computed with several values of abundance was applied. In this way, the elemental abundance was fit to the disk-center spectra and then fixed to this value for the other $\mu$-angles. Finally, we computed the line profiles at the other $\mu$-angles with the best-fit model and from those we calculated the $W_\lambda$ of the synthetic spectra by direct integration of the line within wavelength ranges that were considered blend-free (see Sect. \ref{sec: EW}). 
For the study of the \ion{Na}{I} D$_1$ and \ion{K}{I} $7699\,\AA$ resonance lines, we also convolved the synthetic spectrum with the CRISP instrumental profile in order to compare $W_\lambda$ with the observed SST data (see Sect. \ref{sec: EW}). 

\subsection{Model atoms}\label{sec: atoms}
Several studies have shown that both \ion{Na}{I} and \ion{K}{I} atoms are affected by strong non-LTE effects (i.e. over-recombination of the ground state), meaning that non-LTE line formation is required to correctly reproduce observations. \citet{Lind2011} has shown that Na stellar abundances can be overestimated by up to 0.5\,dex if calculations in 1D LTE are performed for saturated lines of metal-poor dwarfs and giants ([Fe/H]=-2). 
In the case of the Sun, they found a difference of about 0.02 dex between the 1D LTE and non-LTE inferred abundance. 
\citet{Reggiani2019} showed that the abundance of K in the Sun is overestimated by about 0.3\,dex in 1D LTE compared to 1D non-LTE.

The non-LTE of any line is generally determined by the change in formation depth, which is given by the departure coefficient of the lower level, and the change in line source function ($S_l$), whose ratio with respect to $B_\nu$ is equal to the ratio between the upper and lower level departure coefficients (neglecting stimulated emission). For the resonance lines (Na D and \ion{K}{I} $7699\,\AA$), it has been seen that $S_l$ resembles that of a pure scattering line in a two-level atom at all depths, meaning that $S_l$ is equal to the profile-averaged mean intensity ($\Bar{J}_\phi$) and strongly subthermal, that is $\Bar{J}_\nu < B_\nu$. Over-recombination and over-deexcitation (photon loss) overpopulate the ground levels of \ion{Na}{I} and \ion{K}{I} compared with LTE, pushing the mean formation depth outwards in the atmosphere. The combined effect is a stronger line in non-LTE compared to LTE, and specifically a deeper line core. For more details about the non-LTE effects of \ion{Na}{I}, we refer the reader to Sect. 3.1 of \citet{Lind2011}, and for \ion{K}{I} to Sect. 4.3 of \citet{Reggiani2019}.

%%%%%% Intro to Sodium %%%%%%
The model atom used in \texttt{Balder} for the Na atom is the one developed and tested in \citet{Lind2011}, characterized by a structure of 23 levels in total: 22 energy levels for \ion{Na}{I} and the ionized sodium (\ion{Na}{II}) continuum. The lines of the Na doublet originate from the transition between the levels 3s-3p$_{1/2}$ and 3s-3p$_{3/2}$. Both fine and hyper-fine structures are accounted for in the model atom. The latter describes both the radiative and collisional transitions among the energy levels and, in particular, detailed quantum mechanical calculations are used for the inelastic collisions of Na respectively with electrons (\citealt{Park1971}; \citealt{Allen1993}; \citealt{Igenbergs2008}) and hydrogen atoms (\citealt{Belyaev2010}; \citealt{Barklem2010}). Regarding the latter, in this work, the impact
of rate coefficients for the collisions with neutral hydrogens from \citet{Kaulakys1991} was
also explored.  Indeed in the model atom developed in \citet{Lind2011}, these rates were only considered above the ionic limit, but \citet{Amarsi2018b, Amarsi2019b} argued that it could be necessary to add these rate coefficients even below the ionic limit, at least when asymptotic models such as those of \citet{Barklem2016b} or \citet{Belyaev2013} are used. 
The results of  \citet{Amarsi2018b, Amarsi2019b} for O and C, as well as \citet{Bergemann2019} for Mn, suggest that doing so leads to better agreement of synthetic spectra with observational data.

%%%%% POTASSIUM ATOM %%%%%
The model atom of the potassium atom was recently developed by \citet{Reggiani2019}, who used the most up-to-date and accurate atomic data, for both the radiative and collisional transitions (e.g., \citealt{Belyaev2017}; \citealt{Yakovleva2018}; \citealt{Barklem2017}). 
The atomic model consists of 19 energy levels, including the ionized potassium (\ion{K}{II}) continuum. 
For the collisions with hydrogen atoms, they adopted rate coefficients from \citet{Yakovleva2018}, based on the asymptotic models of \citet{Barklem2016b}. To these data, they also added those from the free electron model of \citet{Kaulakys1991}, following \citet{Amarsi2018b, Amarsi2019b}. From \citet{Reggiani2019}, the K+H collisions significantly affect the abundance determination. In particular, they produce an abundance difference of about 0.03 dex in the Sun if not taken into account.
In the same work, they have shown that the resonance line at $7699\,\AA$ has wings insensitive to departures from LTE, but a core that is strongly affected by non-LTE, as it is much deeper than in LTE.

Both these model atoms were employed for full 3D non-LTE calculations in a recent paper by \citet{Asplund2021} to study and update the chemical composition of the Sun. However, the lines selected to compute abundances are not the same as those analyzed in this work. The atomic data of the Na lines and \ion{K}{I} resonance line studied here are summarised in Table \ref{tab: atomicdata}.

%%%%%%%%%%%%%%%%%%%%%%%%%%%%%%%%%%%%%%%%%%%%%
%----------------------------------------------------------%
%----------------ANALYSIS--------------%
\section{Analysis of the results}\label{sec: analysis}
The CLV of solar lines represents a crucial test for line formation modeling since it is an excellent probe of the atmospheric structure. Indeed, the light emitted at different $\mu$-angles probe different heights of the solar atmosphere: shallower layers close to the limb, and deeper layers at the disk center ($\mu=1.0$). 
3D radiative transfer shows that the profiles of the lines emitted from different regions of the solar surface vary in both shape and strength. This is important for transmission spectroscopy studies when the stellar lines are distorted due to the planet transiting in front of its star, and blocking the light from different regions of the stellar surface. The deformation of the stellar lines often complicates the interpretation of (chemical) spectral signatures caused by the planet's atmosphere, even leading to the detection of false positives for some chemical elements or to the wrong derived abundance in some cases. Therefore high-resolution ground-based spectra need to be corrected for the CLV signature, using a model that is as realistic as possible.

For this purpose, in Sect. \ref{sec: EW} and \ref{sec: shapes} we compare the spatially resolved solar observations of the IAG Atlas and of the SST data with several theoretical models assuming different model atmospheres (1D and 3D) and line formation (LTE and non-LTE). 
Furthermore, in Sect. \ref{sec: sun-jup} we simulate the impact on transit phase curves for the Sun when the CLV is modeled with different assumptions in a Sun-Jupiter system.

\subsection{Center-to-limb variation of equivalent widths}\label{sec: EW}
%%%%%% FIGURES %%%%%
%%%%%%  Na I lines: EW vs mu %%%%%%%
   \begin{figure*}
   \resizebox{\hsize}{!}
            {\includegraphics[width=10cm]{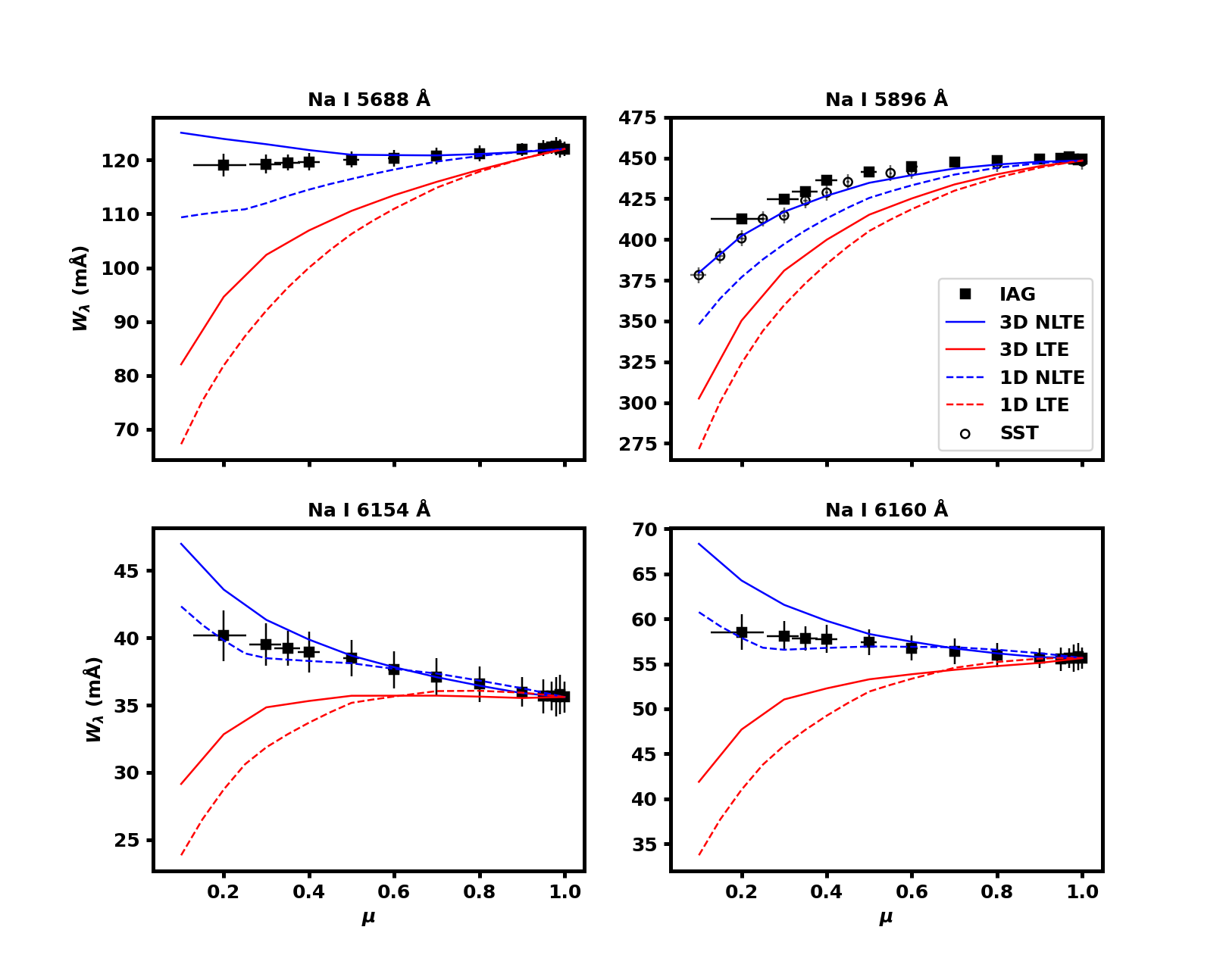}}
            
      \caption{Center-to-limb variation of the solar \ion{Na}{I} lines. Comparison of the equivalent widths of the IAG Atlas (filled squares) and the SST data (open circles) with several different model atmospheres and atomic data, colored as in the figure's legend.} % EW_Na_lines_extraH
              
         \label{fig: EWvsmuNaI}
   \end{figure*}
%%%%%%%%%%%%%%%%%%%%%%%%%%%%%%%%%%%%%
%%%%%%%%%%%% K I EW vs mu %%%%%%%%%%%
\begin{figure}
 \centering
   \includegraphics[width=9cm]{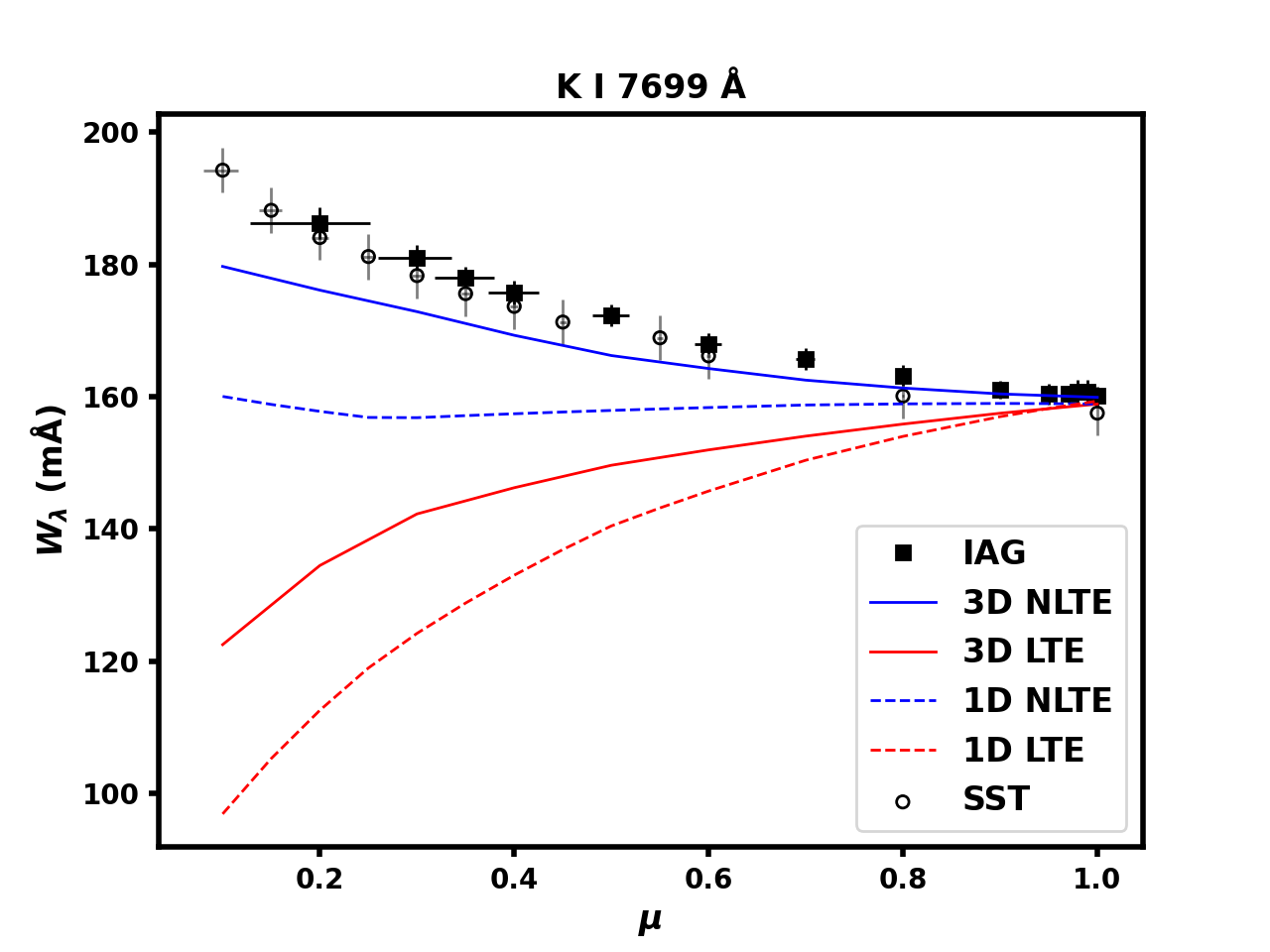}
      \caption{Center-to-limb variation of the \ion{K}{I} 7699\,$\AA$ line. Comparison of the IAG Atlas (filled squares) and the SST data (open circles) with several different model atmospheres and atomic data, colored as in the figure's legend.}
         \label{fig: EWvsmu7699}
   \end{figure}
%%%%%%%%%%%%%%%%%%%%%%%%%%%
%------------------------------------% 
In this section, we analyze the CLV of the equivalent widths of four \ion{Na}{I} lines and one \ion{K}{I} line, presented in Table \ref{tab: atomicdata}. In transmission spectroscopy studies only the strong \ion{Na}{I} D lines are used for the detection of sodium in the planetary atmospheres since they are the only lines strong enough to show a planetary signature. The \ion{Na}{I} D$_2$ line at $5890\,\AA$ in the IAG Atlas was contaminated by too many telluric lines that were cut out from the dataset, hence it was not possible to compare it with our synthetic spectra. However, we analysed the CLV of 3 other \ion{Na}{I} optical lines, respectively at $5688\,\AA$,  $6154\,\AA$ and $6160\,\AA$, in order to further test our models. The \ion{Na}{I} doublet at $6154\,\AA$ and $6160\,\AA$ is also commonly used for abundance analyses of solar-metallicity stars. The CLV of these two lines was studied by computing the $W_\lambda$ via direct integration of the line profile over a wavelength region extending $\pm 0.12\,\AA$ from the line center, in order to exclude nearby lines from other elements. For the \ion{Na}{I} $5688\,\AA$ line a wavelength range of $\pm 0.15\,\AA$ was adopted instead. The wavelength range for integration is shown in the last column of Table \ref{tab: atomicdata}. For the \ion{Na}{I} D$_1$ and \ion{K}{I} resonance lines, the same wavelength range used in Sect. \ref{sec: comparison} was chosen in order to compare the models with both SST and IAG data. For this reason, the convolution of CRISP instrumental profile was applied to the synthetic spectra of these two lines. In all comparisons, the abundance is adjusted to fit the line profile at disk center, as mentioned in Sect. \ref{sec: model}. The resulting $W_\lambda$ as a function of $\mu$-angles for the \ion{Na}{I} lines and the \ion{K}{I} line are shown in Fig. \ref{fig: EWvsmuNaI} and \ref{fig: EWvsmu7699}, respectively.

From both these figures, it is clear that 1D LTE modeling (dashed red line) heavily fails in describing the CLV of the solar lines, underestimating the line strengths at the limb as much as 50.1\% for the \ion{K}{I} and 28.2\% for \ion{Na}{I} D$_1$ at $\mu=0.1$, in comparison with the SST data. For the other \ion{Na}{I} lines compared with the IAG data at $\mu=0.2$, the underestimate is as large as 31\% for $5688\,\AA$ and about 29\% for the 6154 and $6160\,\AA$ lines. Using a 3D atmosphere with LTE line formation (solid red line) slightly improves the CLV but it still considerably under-performs at the limb, with an equivalent width difference of about 20\% for the Na lines, and 37\% for the K line.

The observations are much better reproduced by the 1D non-LTE model (dashed blue line), which is in agreement with the \ion{Na}{I} D$_1$ SST data at $\mu=0.1$ within about 8\%. This demonstrates that the non-LTE effect is a very important feature that must be taken into account in order to accurately model the spectral lines. The agreement with the data improves also for the other Na lines, down to 7.2\% for the $5688\,\AA$ and about 1\% for the 6154 and $6160\,\AA$ lines, at $\mu=0.2$. The equivalent width of the 1D non-LTE model is 17.6\% lower than the SST data at $\mu=0.1$ for the \ion{K}{I} resonance line. 

A large improvement is obtained by the 3D non-LTE modeling (solid blue line) at the limb ($\mu=0.2$) for the \ion{K}{I} (6\%), the $5688\,\AA$ (4\%) line, and the \ion{Na}{I} D$_1$ line (2.5\%). The CLV of the latter is also in excellent agreement with the SST data, with a difference of only 0.4\% at $\mu=0.1$. Nevertheless, the 3D non-LTE model slightly overestimates the strength of the other \ion{Na}{I} lines by about 8.5\% ($6154\,\AA$) and 9.8\% ($6160\,\AA$); that is, performing slightly worse than 1D non-LTE. 
We stress that the slightly better agreement in 1D non-LTE is likely coincidental, since it is dependent on the various parameters in the 1D models that can be tuned to reproduce observations, such as $v_\mathrm{mic}$ and $v_\mathrm{mac}$. The imperfect agreement of the 3D non-LTE model close to the limb can be explained by several uncertain factors in the modeling, such as the upper layers of the \texttt{Stagger} model atmosphere, as well as the rates for collisions with H and electrons in the model atom.

However, it is worth noting that in general the 3D non-LTE prediction performs extremely well given that it does not rely on any free parameter. Considering that for other stars we cannot take spatially resolved observation of their disk, we would not be able to constrain the parameters for the 1D modeling, like we can do for the Sun.
 
Moreover, the CLV of the continuum and other observational diagnostics are not well reproduced by 1D models, such as \texttt{MARCS}, not even in the Sun, as shown in \citet{Pereira2013}.  
For other stars, we cannot resolve the stellar disk, and therefore we cannot take observations at different $\mu$-angles to fit the free parameters of our 1D models with. For this reason, it is important to note how well we can reproduce the solar CLV without any free parameter with our 3D non-LTE models.

\subsection{Center-to-limb variation of line profiles}\label{sec: shapes}
%%%%%%%%%%%% FIGURES AND TABLES %%%%%%%%%%%%
%%%% Line profile of Na D1 in ALL models (1D vs 3D NLTE) %%%% the others are in appendix!
% From plot_bestfit2.ipynb
\begin{figure*}
 \centering
   \resizebox{\hsize}{!}{\includegraphics[width=9cm]{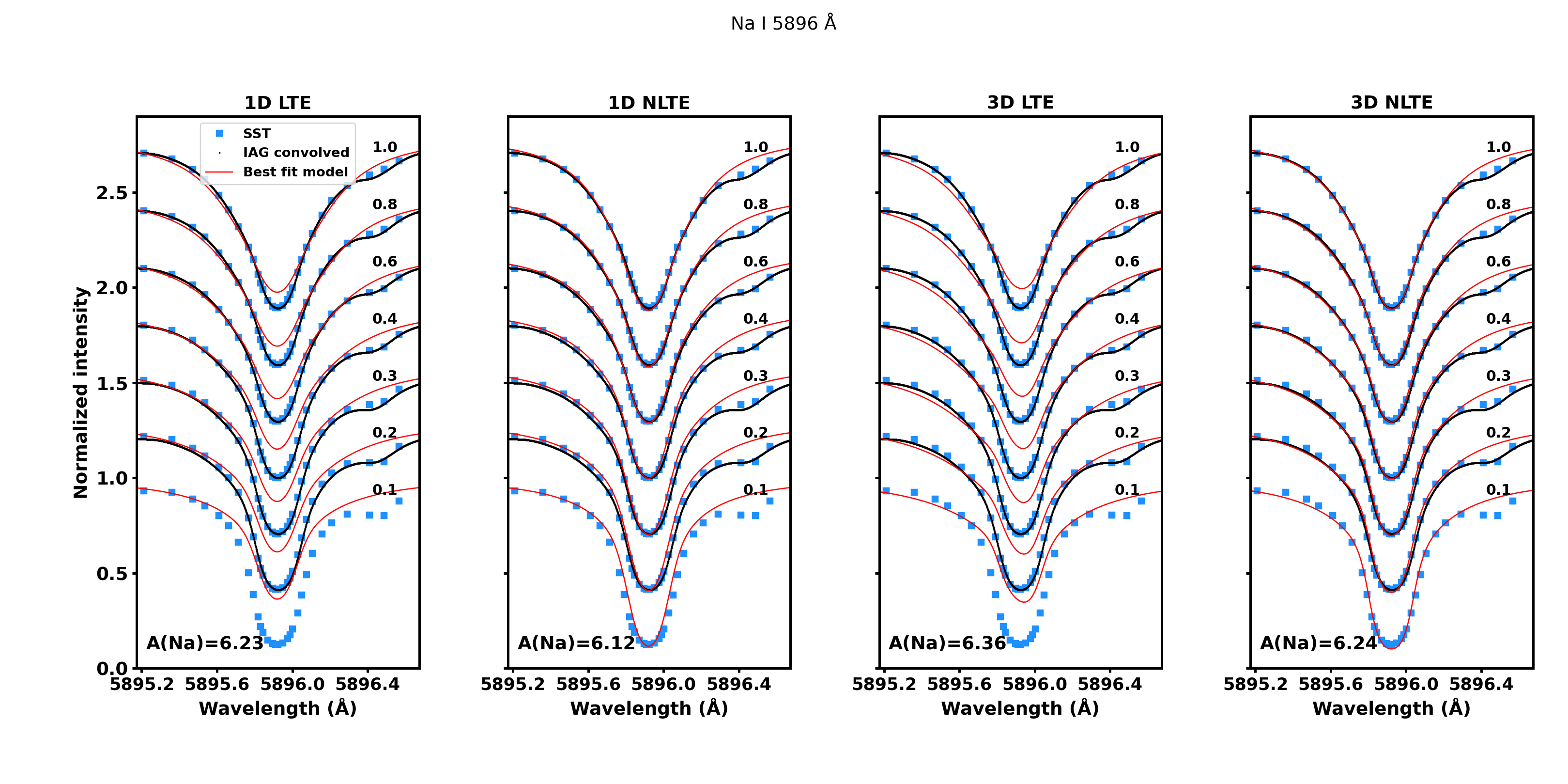}}%
      \caption{Normalized observed (black dots and blue squares) intensity and synthetic (red lines) center-to-limb profiles for the \ion{Na}{I} D$_1$ line, where the numbers above each spectrum correspond to the $\mu$-angle. In the left corner, the calibrated abundance is shown. Spectra for $\mu \geq 0.2 $ have been incrementally offset vertically by $+0.3$ for clarity.
      }
         \label{fig: line5896}
   \end{figure*}
%%%%% Line profile of K I in ALL models (1D vs 3D NLTE) %%%%
\begin{figure*}
 \centering
   \resizebox{\hsize}{!}{\includegraphics[width=9cm]{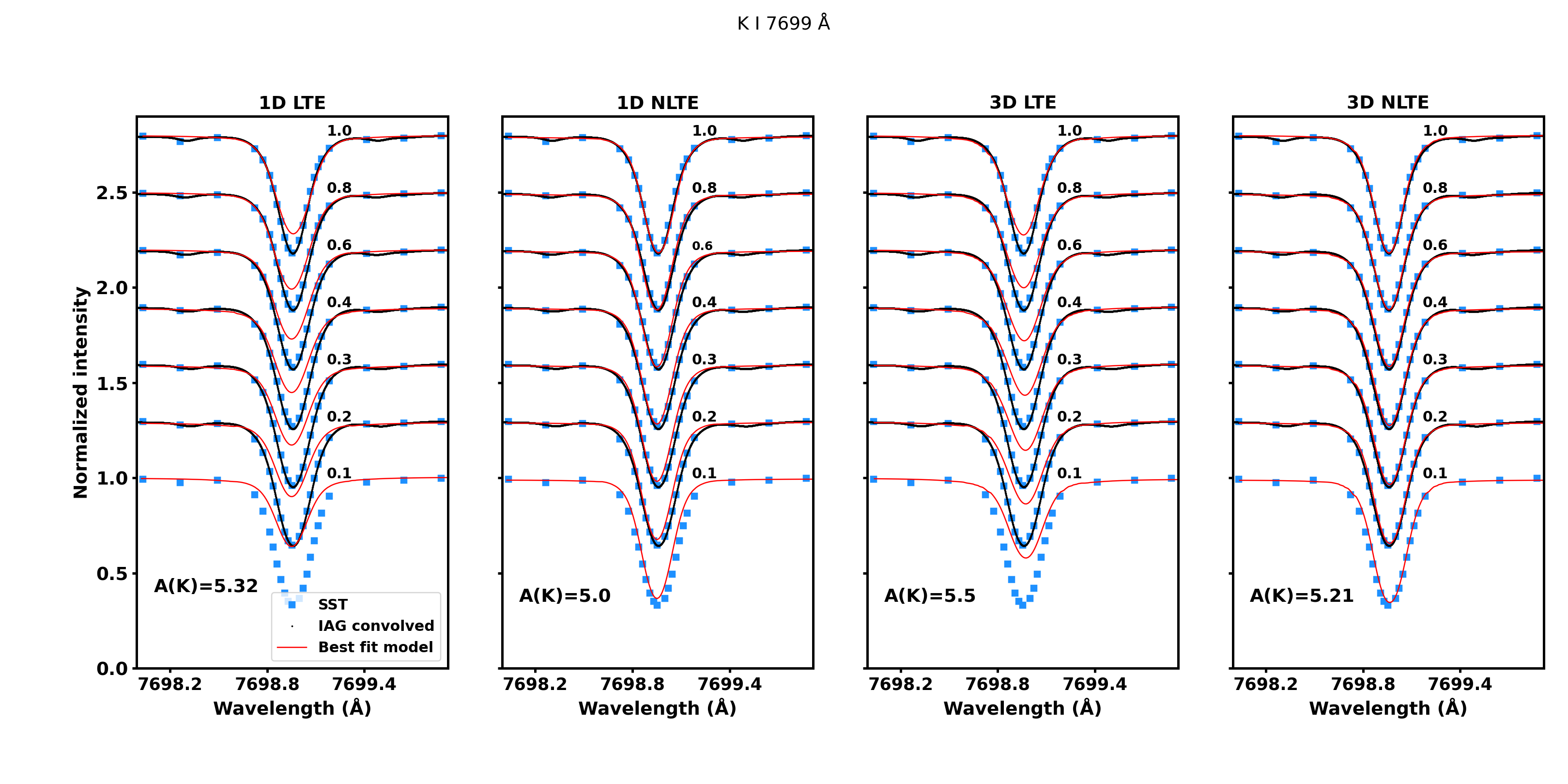}}
      \caption{Normalized observed (black dots and blue squares) intensity and synthetic (red lines) center-to-limb profiles for the \ion{K}{I} $7699\,\AA$ line, where the numbers above each spectrum correspond to the $\mu$-angle. In the left corner, the calibrated abundance is shown. Spectra for $\mu \geq 0.2 $ have been incrementally offset vertically by $+0.3$ for clarity.} 
         \label{fig: line7699}
   \end{figure*}
%%%%%%%%%%%%%%%%%%%%%%%%%%%%%%%%%%%%%%%%%%%%%%%%%%%%%%%%%%%%%%%%%%%%%%%
%%% Table with the derived Abundance %%%
\begin{table}
\centering 
\caption{Solar abundances determined from the best-fitting synthetic spectrum at disk center for four \ion{Na}{I} lines and one \ion{K}{I} line.}\label{tab: abundance} 
\resizebox{\columnwidth}{!}{\begin{tabular}{c c c c c} 
\\ \hline \hline 
$\lambda$ (\AA)  & 1D LTE & 1D non-LTE & 3D LTE & 3D non-LTE \\ \hline
\ion{Na}{I} & & & & \\
5896 & 6.23 & 6.12 &  6.36 & 6.24\\ %6.08 in 1D non-LTE, 6.20 in 3D non-LTE  
5688 & 6.21 & 6.08 & 6.35 & 6.19 \\ %6.18 3D non-LTE 
6154 & 6.26 & 6.20 & 6.31 & 6.26 \\
6160 & 6.26 & 6.19 & 6.34 & 6.26 \\ 
$\left\langle A(\mathrm{Na})\right\rangle$ & $6.24 \pm 0.02$ & $6.15 \pm 0.05$ & $6.34 \pm 0.02$ & $6.23 \pm 0.03$\\ \hline
\ion{K}{I} & & & & \\
7699 & 5.31 & 5.00 & 5.49 & 5.21 \\ \hline %add errors?
\end{tabular}}\\
\end{table}
%%%%%%%%%%%%%%%%%%%%%%%%%%%%%%%%%%%%%%%%%%%%%%%%%%%%%%%%%%%%%%%%%%%%%%%
%-----------------------------------------%
Although for some \ion{Na}{I} lines the 1D non-LTE modeling is predicting a very good CLV in terms of equivalent widths (Sect.~\ref{sec: EW}), the same is not true anymore for what regards the line profiles. The latter are shown in Fig. \ref{fig: line5896} and \ref{fig: line7699} for the \ion{Na}{I} D$_1$ and \ion{K}{I} line, respectively, displaying the intensity of the $\mu$-pointings in common between the IAG and the SST data, namely $\mu= 0.2, 0.3, 0.4, 0.6, 0.8, 1.0$, and the $\mu=0.1$ profile in comparison with the SST profile. The line profiles of the other Na lines compared with the IAG data only, are shown in Appendix in Fig. \ref{fig: line5688}, \ref{fig: line6154} and \ref{fig: line6160}. Moreover, Fig. \ref{fig: line5896IAG} and \ref{fig: line7699IAG} show the line profiles of the \ion{Na}{I} D$_1$ and \ion{K}{I} $7699\,\AA$ of the IAG data at original resolution compared with the non-convolved synthetic spectra.
In all the figures, from left to right, the panels display the data with the best-fit 1D LTE, 1D non-LTE, 3D LTE and 3D non-LTE synthetic spectra overplotted. 

As is apparent from all these figures, the 3D non-LTE modeling is able to consistently reproduce the observed line profiles from the disk center to the limb. On the other hand, the shape and asymmetries of the line profiles are not well modeled by the 1D non-LTE spectra, getting worse at the limb even if the disk center intensity matches the data quite well. 
The LTE spectra instead fail to correctly reproduce the shape of the lines even at disk center. 
The $W_\lambda$ of the observations at different $\mu$-angles are reported in Table \ref{tab: EWSST} and \ref{tab: EWIAG} in Appendix for the SST and IAG data, respectively. The best-fitting abundances are reported in Table \ref{tab: abundance}, from which we notice that the average Na abundance from 3D non-LTE models is in very good agreement with the value of $A(\mathrm{Na})=6.22 \pm 0.03$, derived by \citet{Asplund2021} even if they used a different line selection. The \ion{K}{I} abundance instead, in this work is higher by about 0.14\,dex compared to the \citet{Asplund2021} value. 
However, it is important to note that the \ion{Na}{I} D and \ion{K}{I} resonance lines are not ideal abundance diagnostics as they are saturated lines. Hence, the values in \citet{Asplund2021} are preferred in the context of the solar chemical composition.

The reason why the 3D non-LTE modeling overestimates the line strengths $W_\lambda$ at the limb in \ion{Na}{I} lines (Fig. \ref{fig: EWvsmuNaI}) is not fully clear and we investigated the impact of collisions with neutral hydrogen (H). 
As discussed in Sect~\ref{sec: atoms}, earlier work has argued that it could be necessary to add rate coefficients from asymptotic models such as those of \citet{Barklem2016b} or \citet{Belyaev2013}, to those from the free electron model of \citet{Kaulakys1991}.  
Motivated by this, the Na model atom tested here includes additional rate coefficients from \citet{Kaulakys1991} for the Na+H collisions, even though the base data are not from an asymptotic model, but come from detailed quantum mechanical calculations.
We found that the additional rates have very little influence on the weak $6154$ and $6160\,\AA$ lines, whereas the effect is more noticeable for the stronger $5688\,\AA$ and $5896\,\AA$ lines. Specifically, for the latter the 3D non-LTE model matches the CLV of the equivalent widths of the SST data much better, as can be seen from Fig. \ref{fig: EWvsmuNaI}. The comparison with the previous model atom developed in \citet{Lind2011} is shown in Appendix in Fig. \ref{fig: atom-comparison}. With the previous model atom, the $W_\lambda$ of the 3D non-LTE model prefers the IAG data, whereas in the new model atom (the one adopted in this work) the 3D non-LTE model is in better agreement with the SST data. 
Apart from uncertainties on the H collisions, other factors that might explain why the 3D non-LTE models do not perfectly match observations of some Na lines and the \ion{K}{I} line at the limb, could be the uncertainties on the upper layers of the 3D model atmosphere as well as the vertical resolution of the model. 

In the next section, we demonstrate the effect of CLV in narrowband transmission spectroscopy with the example of a Jupiter-sized planet orbiting around a Sun-like star. For this purpose, the best-fitting synthetic spectra are used to evaluate the CLV effect in a simulated Sun-Jupiter system in order to analyze the impact of different models and different assumptions on the measurement of planetary atmospheric species. Also, the CLV impact on different geometries of the planet-star system and on different lines is investigated.

\subsection{Sun-Jupiter system}\label{sec: sun-jup}
%%%%%%%%%%%%%%%% FIGURES AND TABLES %%%%%%%%%%%%%%%%%%%%
%%%% TABLE WITH SUMMARY OF SIMULATIONS RUN ON StarRotator %%%%
\begin{table}
\centering 
\caption{Input parameters for the simulated Sun-Jupiter system.}\label{tab: Jup-setup} 

\begin{tabular}{l c} \hline \hline 
\noalign{\smallskip}
Parameters & Value \\ \hline
$a/R_\odot$ $^{(a)}$ & 1117.9 \\ %scaled semi-major axis 
$e$ $^{(a)}$ & 0.0487 \\ %eccentricity (actually 0.0)
$\omega$ & 0.0 \\ %argument of pericenter 
Obliquity & 0.0 \\
$b=\frac{a}{R_\odot} \cdot \cos(i)$ & 0.0, 0.2, 0.4, 0.6, 0.8, 1.0 \\
$R_\mathrm{p}/R_\odot$ $^{(a)}$ & 0.10039 \\
$P_\mathrm{orb}$(days) $^{(a)}$ & 4380 \\
mode & phases \\ 
$v_\mathrm{rot}$(m s$^{-1}$) & 0, 2069.44$^{(b)}$, 10000$^{(c)}$\\ \hline 
\noalign{\smallskip}
\end{tabular}\\
\raggedright 
\textbf{Notes.}\\
    $^{(a)}$ From the Jupiter Fact Sheet from NASA\tablefootnote[10]{\url{https://nssdc.gsfc.nasa.gov/planetary/factsheet/jupiterfact.html}};\\
    $^{(b)}$ From \citet{Meftah2014}, to simulate a slow-rotating Sun-like star;\\
    $^{(c)}$ to simulate a fast-rotating star.
\end{table}
%%%%%%%%%%%%%%%%%%%%%%%%%%%%%%%%%%%%%%
%%%% Figure with different simulated b %%%%
\begin{figure}
    \centering
    \includegraphics[width=7cm]{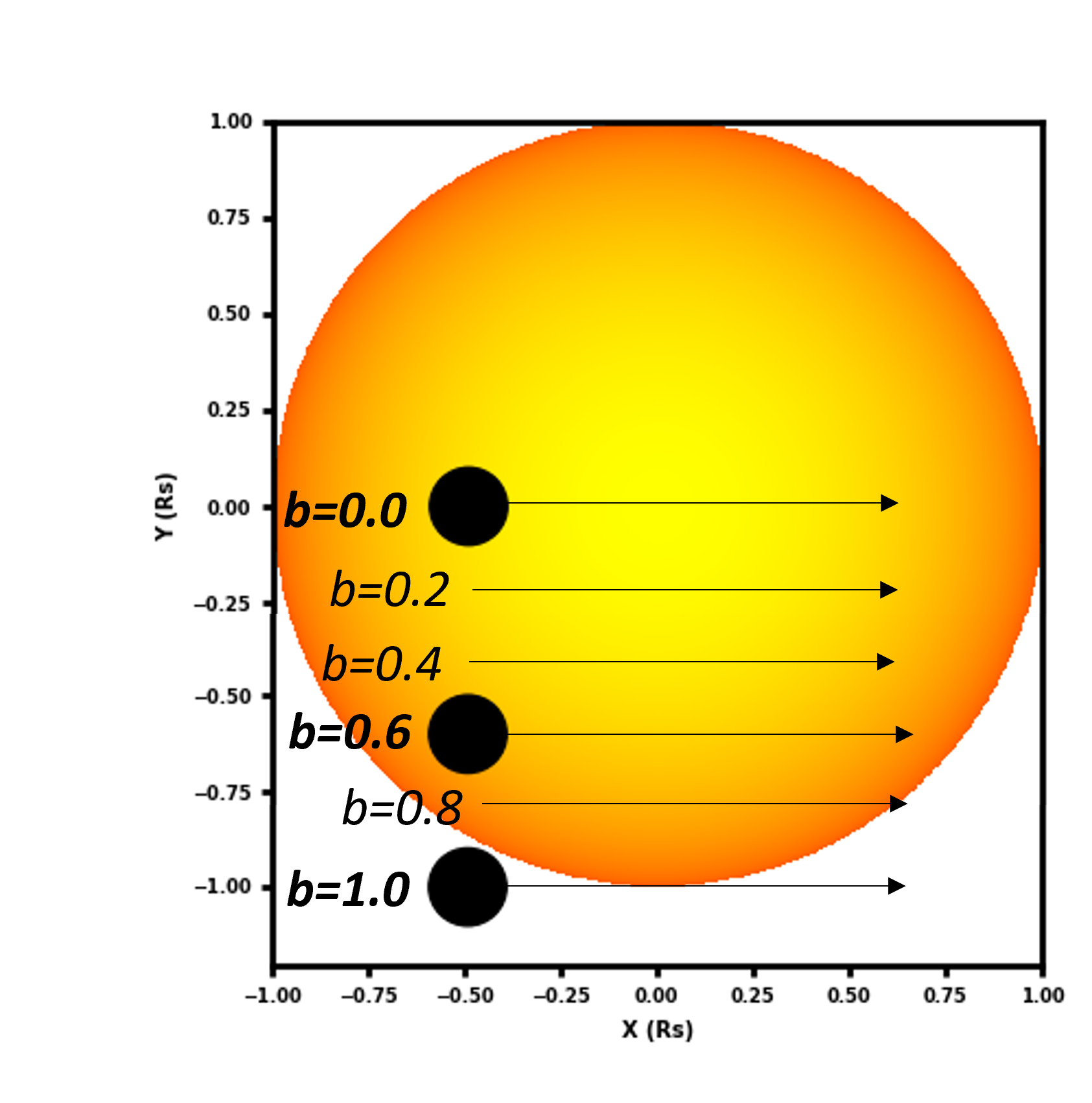}
    \caption{Transit schematic with different impact parameters.}
    \label{fig: Jup b}
\end{figure}
%%%%%%%%%%%%%%%%%%%%%%%%%%%%%%%%%%%%%%%%%%%%%%%%%%%
\begin{figure} 
\centering
\includegraphics[width=7cm]{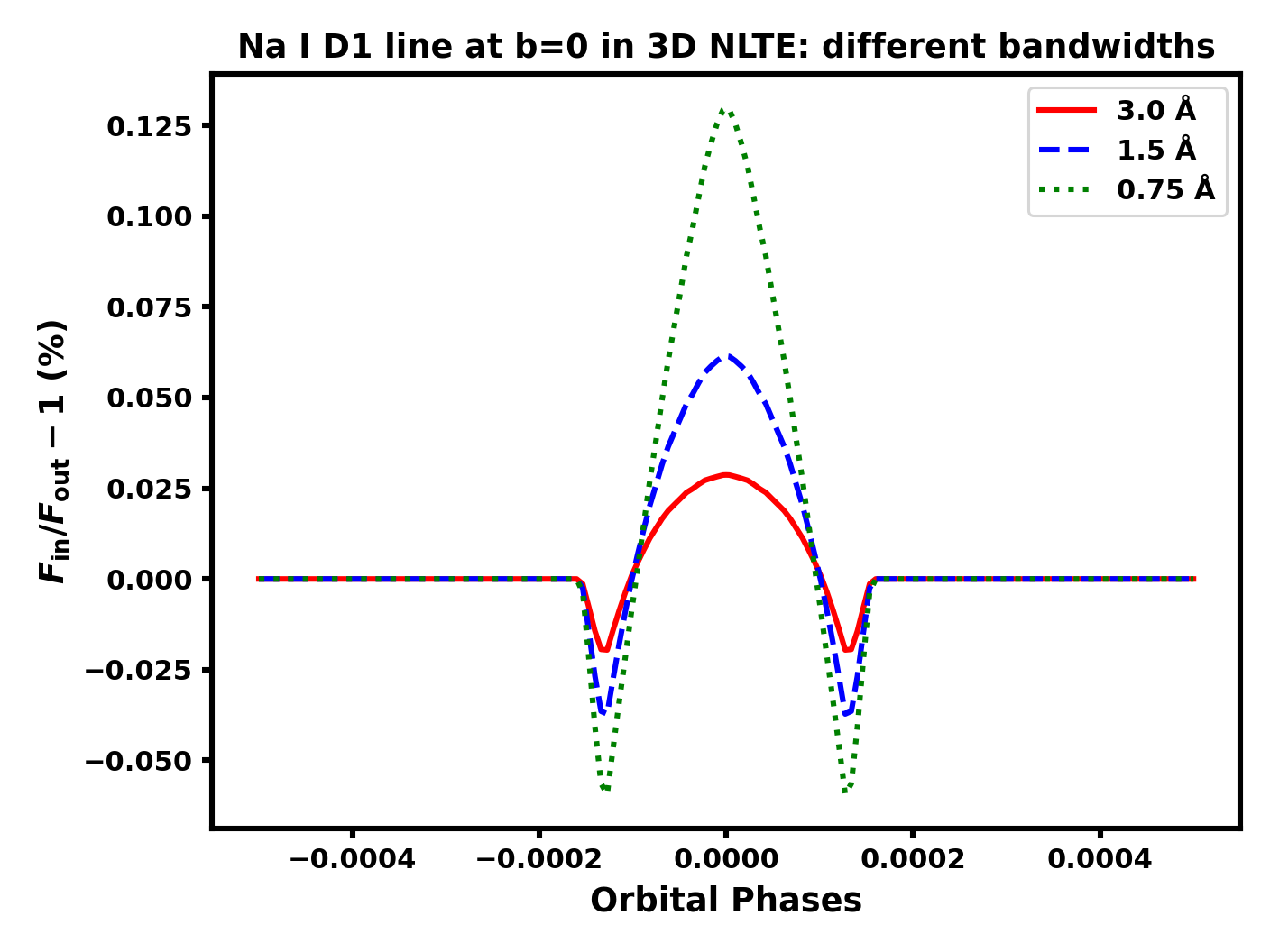}\hfil \centering
\caption{Transmission curves of the \ion{Na}{I} 5896 $\AA\,$ line for a Jupiter-sized planet transiting a Sun-like star in an edge-on transit (i.e., $b=0$), evaluated in different bandwidths around the line center, colored as in legend. 3D non-LTE synthetic spectra are used.}\label{fig: CLV3DNLTE_diffbands}
\centering
\end{figure}
%%%%%%%%%%%%%%%%%%%%%%%%%%%%%%%%%%%%%%%%%%%%%%%%%%%%
%%%%%%%%%% CLV models in 3D NLTE for DIFFERENT b in 0.75 Å %%%%%%%%% 
\begin{figure} 
\centering
\includegraphics[width=7cm]{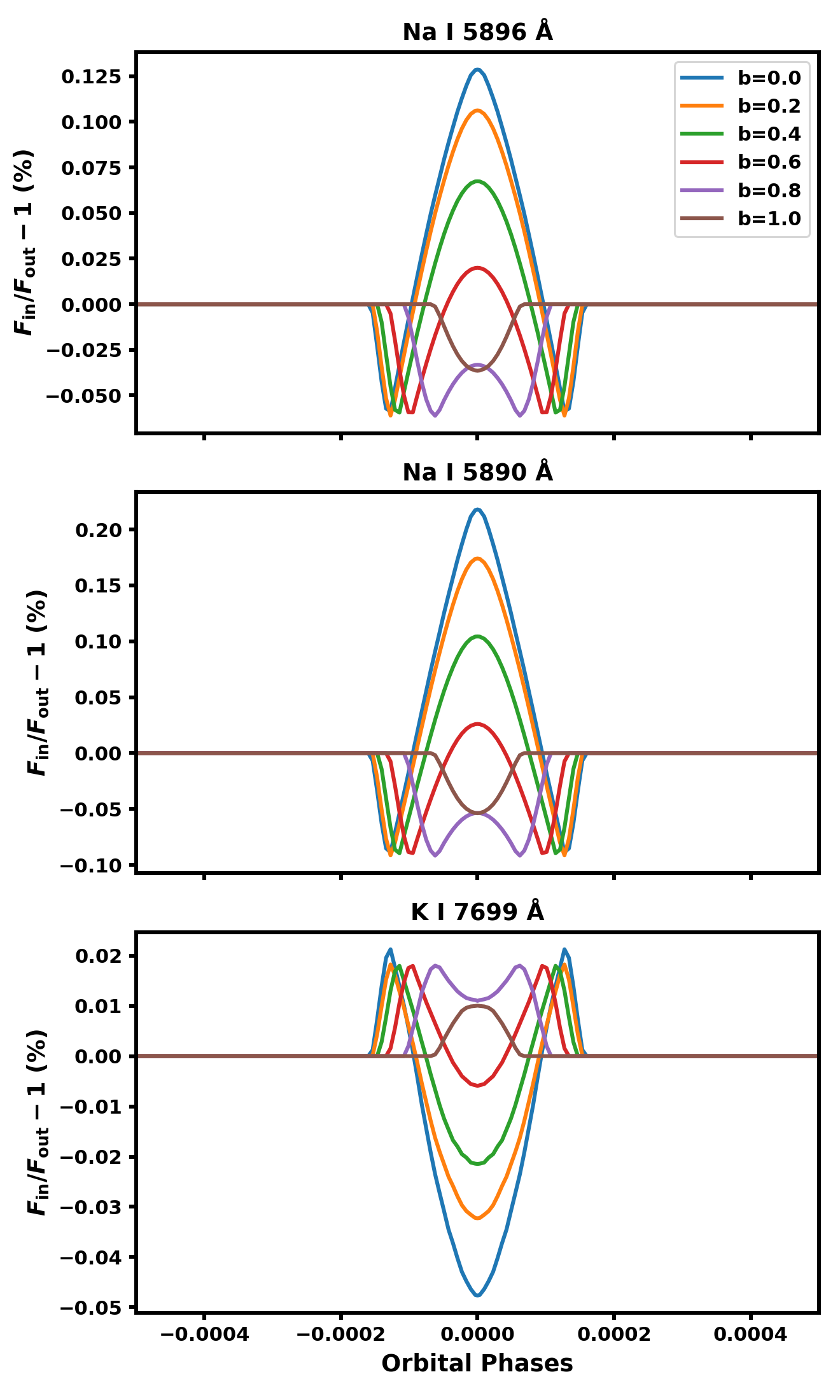}\hfil \centering
%CLVcurves_alllines_v0.png
\caption{Transmission curves of the \ion{Na}{I} D$_1$ (top), \ion{Na}{I} D$_2$ (middle) and \ion{K}{I} line (bottom) for a Jupiter-sized planet transiting a Sun-like star in different trajectories, colored as in legend. The CLV curves are made using synthetic spectra in 3D non-LTE. The relative flux is computed in a bandwidth of 0.75\,$\AA$ on the line center.}\label{fig: CLV3DNLTE_diffb}
\centering
\end{figure}
%%%%%%%%%%%%%%%%%%%%%%%%%%%%%%%%%%%%%%%%%%%%%%%%%%%%
%%%%%%%%%%%% CLV curves at b=0 for DIFFERENT MODELS and SST data in 0.75 Å %%%%%%%%% 
\begin{figure}
\centering
\includegraphics[width=9cm]{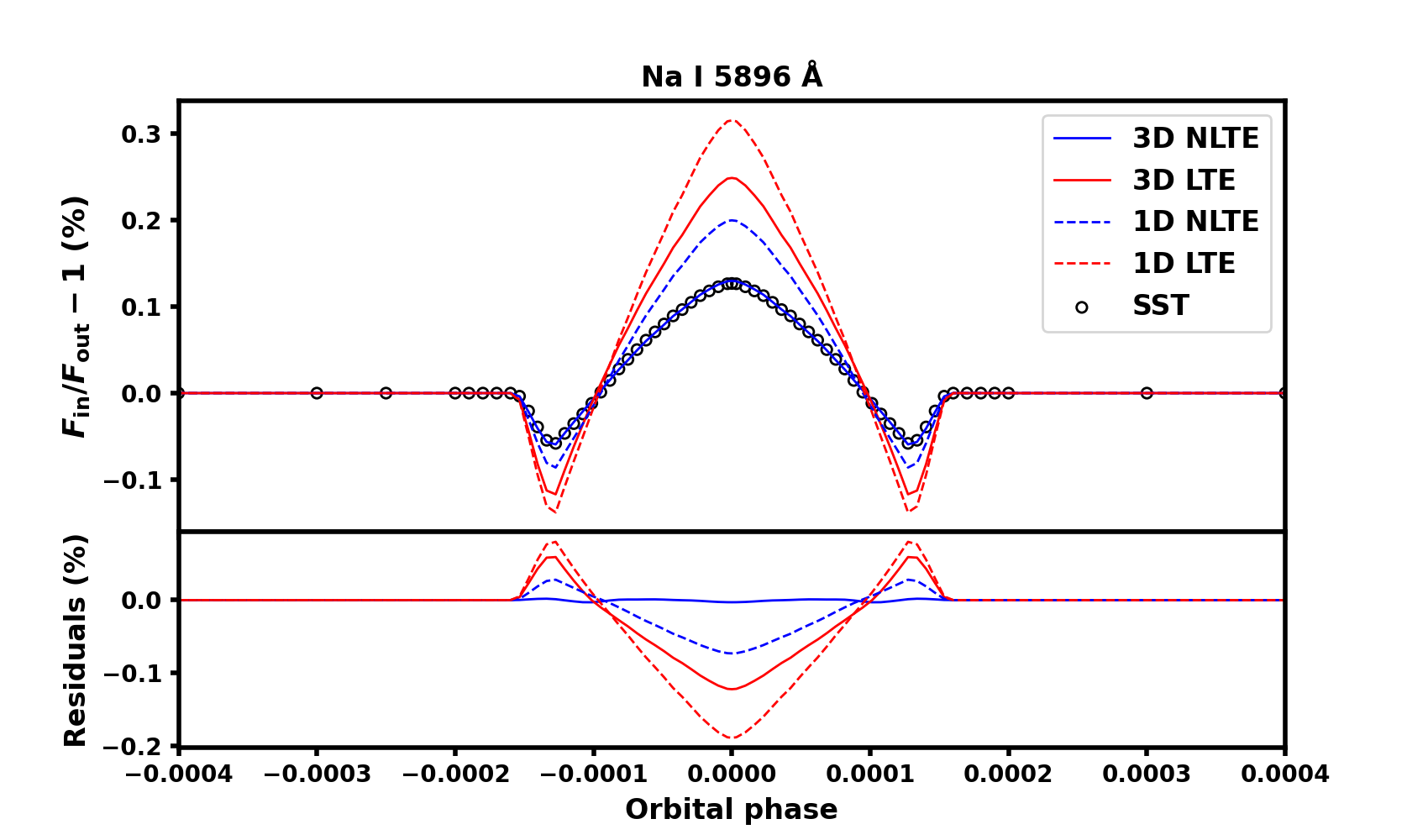}\hfil \centering
\includegraphics[width=9cm]{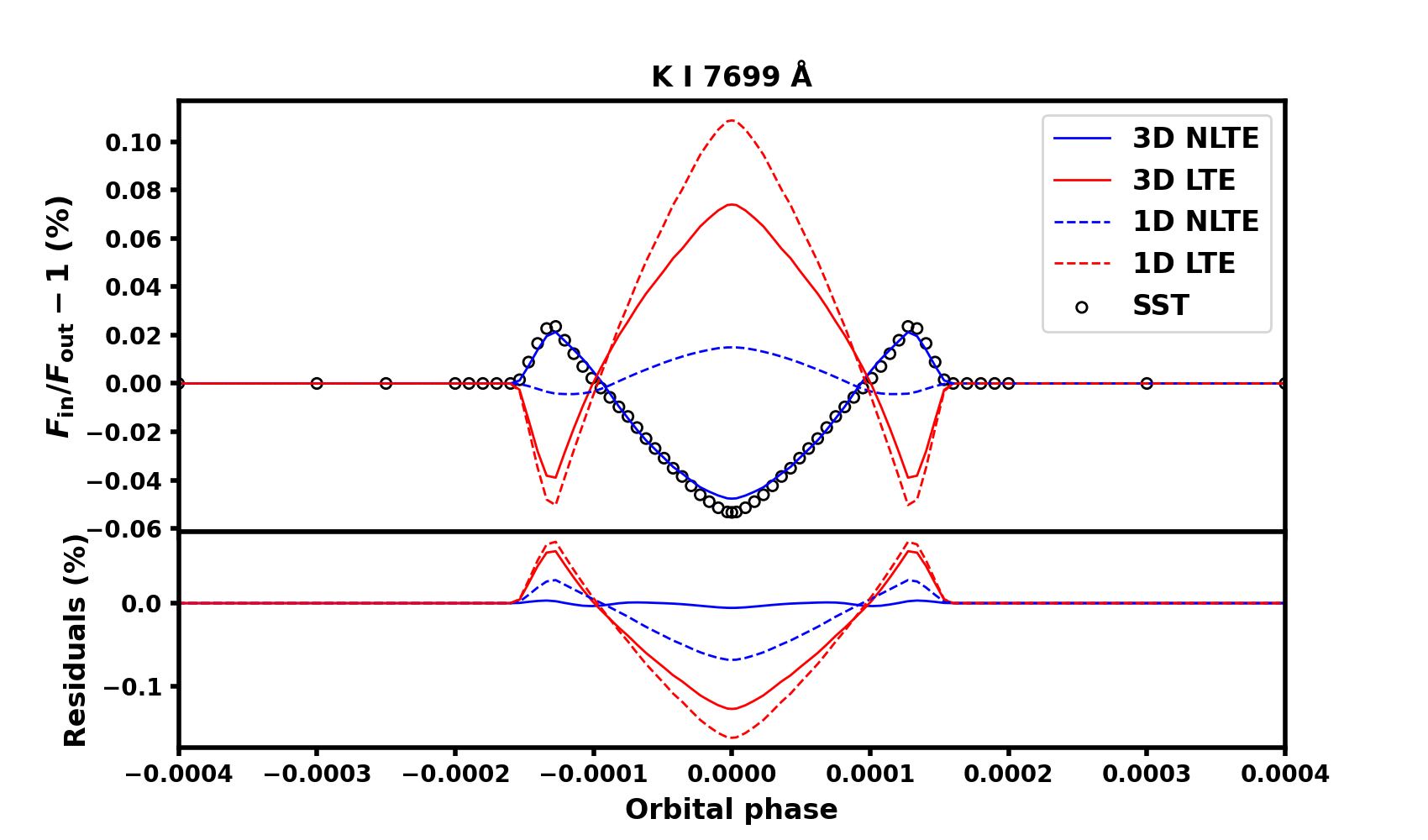}\hfil \centering
\caption{Transmission curves and residuals of the \ion{Na}{I} D$_1$ (top) and \ion{K}{I} line (bottom) for a Jupiter-sized planet transiting a Sun-like star at $b=0$, using different models and the SST data to compute the stellar spectrum, as in legend. The relative flux is computed in a bandwidth of 0.75 $\AA\,$ around the line center. The residuals are the difference between the curve given by the SST data and the models in percentage.}\label{fig: CLV_diffmodels}
\centering
\end{figure}
%%%%%%%%%%%%%%%%%%% RM figures %%%%%%%%%%%%%%%%%%%%% 
\begin{figure}
\centering
\includegraphics[width=9cm]{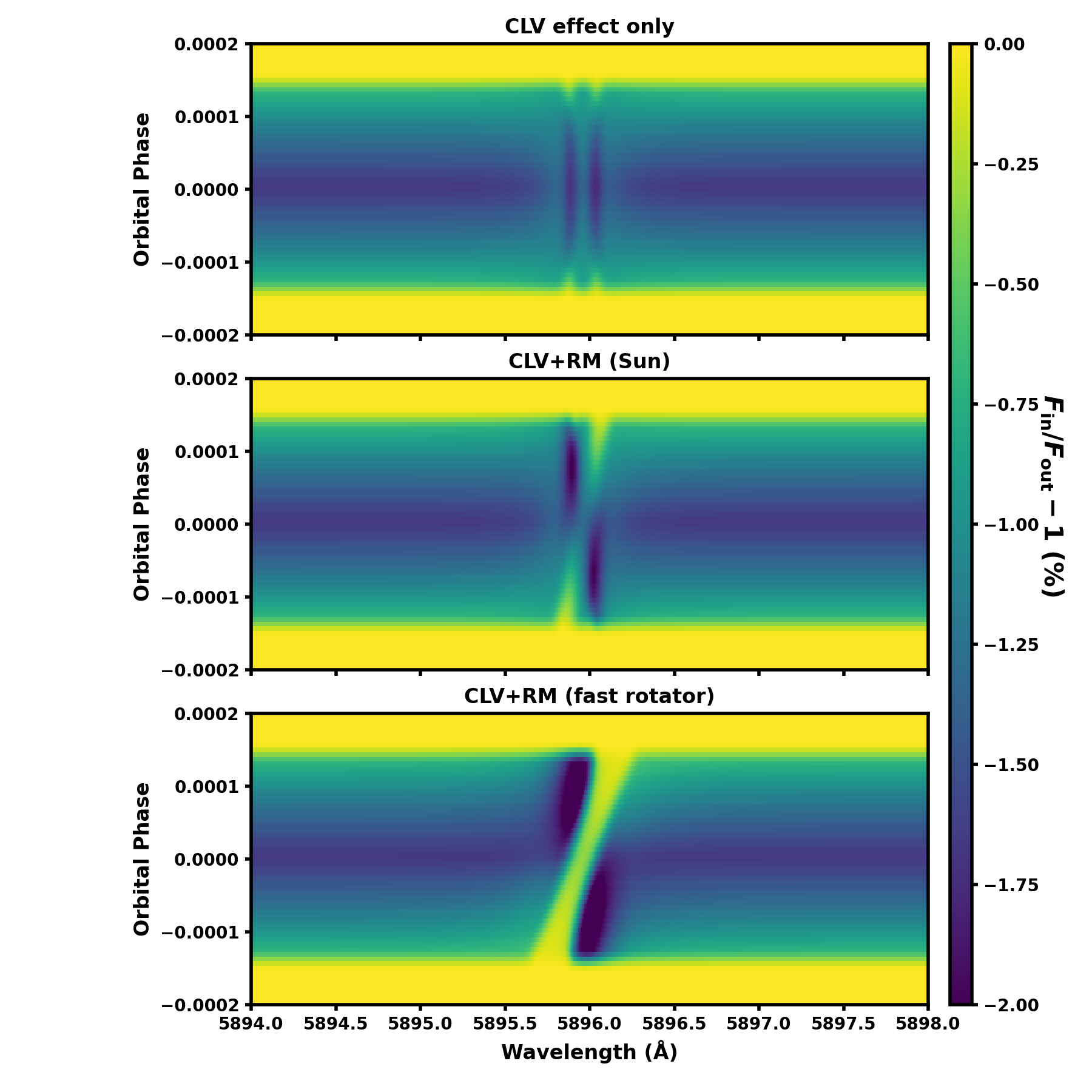}\hfil \centering
\caption{Models of the CLV and RM effect of the \ion{Na}{I} D$_1$ line for a Sun-like star orbited by a Jupiter-sized planet at $b=0$. The upper panel is the CLV effect-only, and the middle panel combines the CLV with the RM effect for a slow-rotator like the Sun. Finally, the bottom panel shows the two effects for a fast-rotator, that is a Sun-like star with a rotational velocity of $10\,\mathrm{km \ s^{-1}}$.}\label{fig: Dopplershadows}
\centering
\end{figure}
%%%%%%%%%%%%%%%%%%%%%%%%%%%%%%%%%%%%%%%%%%%%%%%%%%%%
\begin{figure} 
\centering
\includegraphics[width=7cm]{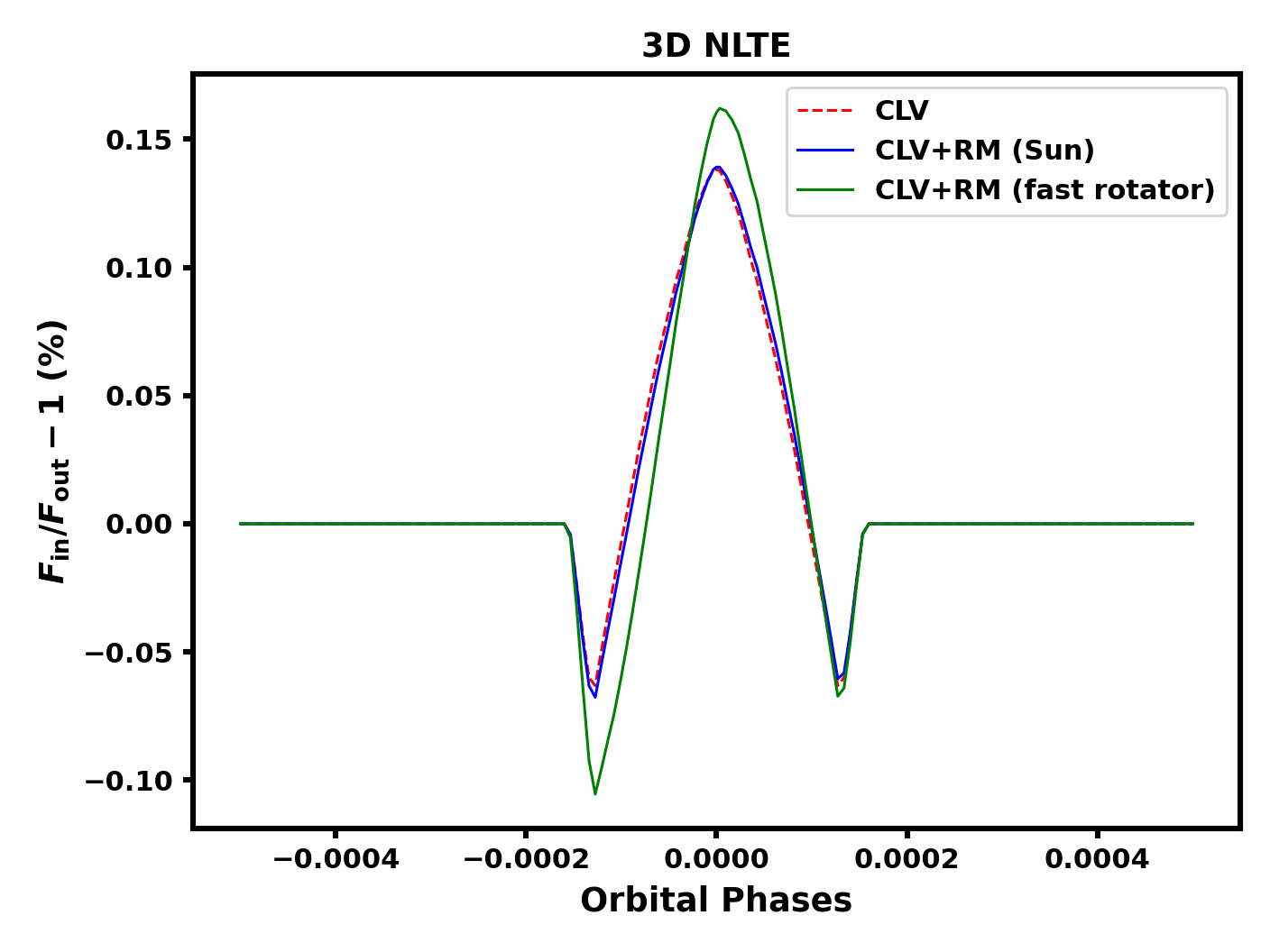}\hfil \centering
\caption{Transmission curves of the \ion{Na}{I} D$_1$ line for a Jupiter-sized planet transiting a Sun-like star in an edge-on transit ($b=0$), for different rotational velocities of the host star. 3D non-LTE synthetic spectra are used and the relative flux is computed in a bandwidth of 0.75 $\AA\,$.}\label{fig: CLV+RM}
\centering
\end{figure}
%%%%%%%%%%%%%%%%%%%%%%%%%%%%%%%%%%%%%%%
%----------------------------------------------------%
The simulation of a transit of a Jupiter-like planet on a circular orbit in front of a Sun-like star is performed by means of a modified version of the \texttt{StarRotator}\footnote[11]{\url{https://github.com/Hoeijmakers/StarRotator}} python package. This code simulates how the rotation-broadened stellar spectrum varies during a planetary transit event. As the transit progresses, the planet blocks the light from different parts of the stellar disk, each contributing differently to the line profile. Consequently, the resulting stellar line profile varies in shape and strength during the transit phases in a non-homogeneous way, also depending on the geometry of the star-planet system (i.e., the pattern of the planet across the stellar disk during the transit). We chose to simulate a circular orbit since most Hot Jupiters are found in low eccentric orbits that have been circularized due to perturbations from nearby stars or tidal forces (e.g., \citealt{Fabrycky2007}). In an eccentric orbit, the transit duration would vary, as well as the ingress/egress shape, while the amplitude of the CLV light curve would not significantly change.
In the simulation, the stellar disk is divided into a grid of (2$\times$400)$\times$(2$\times$400) cells (i.e., 800$\times$800 square pixels), which provides sufficient spatial resolution for our simulations.
Indeed, other works in literature have usually adopted a lower resolution of the grid of (2$\times$100)$\times$(2$\times$100) cells (e.g., \citealt{Wyttenbach2015}; \citealt{Yan2017}). But, we performed a simulation with a lower resolution (i.e., (2$\times$100)$\times$(2$\times$100)), finding a maximum difference in the resulting spectra of about 0.01\%, which is non-negligible for our CLV study. We also tested a higher resolution grid with $(2\times800) \times (2\times800)$ elements and we found a maximum difference of only 0.001\% among the spectra, which instead is insignificant for our purposes. Therefore, we decided to run our simulations with a (2$\times$400)$\times$(2$\times$400) grid which shows a good balance of resolution and computational time.

Each of the elements of the stellar surface is characterized by its own properties, in terms of $\mu$-angle, corresponding radial velocity ($v_\mu$), and intensity ($I_\mu$). The latter comes from the best-fit spectra synthesized with \texttt{Balder} in 3D non-LTE, linearly interpolated at 21 positions between the disk center and the edge, with a spacing of $\Delta\mu=0.05$. Other studies employed the same number of spectra at different angles (e.g., \citealt{Czesla2015}; \citealt{Sicilia2022}). In order to avoid numerical problems at the edge of the stellar disk, the final angle was chosen to be $\mu=0.001$ instead of 0. For the cells at intermediate $\mu$-angles, a linear interpolation between the spectra of the grid is performed. 
This is the main difference with the original \texttt{StarRotator} package, in which the spectra describing the stellar atmospheres are modeled with either the \texttt{PHOENIX} code (\citealt{Husser2013}), which does not take into account the CLV of the lines, or \texttt{SPECTRUM} (\citealt{Szente2019}), which considers the CLV effect (i.e., the intensity is given at different $\mu$-angles) but its RT computation is performed in 1D LTE. In the following simulation instead, the stellar spectra are computed in the more realistic 3D non-LTE assumption.

It is important to notice that the planet is only considered as an opaque body without any atmosphere.
In the simulation, the stellar flux is computed by summing up the spectra of the regions of the discretized stellar disk that are not obscured by the planet (i.e., the "unobscured" stellar spectrum, $F_\mathrm{s}$), so that the observed line profile varies at different orbital phases. This translates into the following equation at $i$-th phase of transit:
\begin{equation}
    F_\mathrm{s} [i]= F_\mathrm{out}-F_\mathrm{p} [i]
    \label{eq: Fs}
\end{equation}
Where $F_\mathrm{out}$ is the out-of-transit flux and $F_\mathrm{p}$ is the integrated flux that is obscured by the planet during the transit at $i$-th orbital phase. Then, by dividing $F_\mathrm{s}$ by its continuum, the normalized spectrum for each given orbital phase is obtained.

%%%%%%%%% NEW PART %%%%%
Finally, the transmission curve or CLV curve is obtained as the relative flux ($F_\mathrm{rel}$), defined as the ratio of the normalized in ($F_\mathrm{in}$) to out-of-transit stellar flux ($F_\mathrm{out}$) integrated into a given bandwidth centered on the line core, as a function of orbital phase:
\begin{equation}
    F_\mathrm{rel}= \frac{F_\mathrm{in}}{F_\mathrm{out}}
    \label{eq: Frel}
\end{equation}

The simulations were performed several times with the same set-up, summarised in Table \ref{tab: Jup-setup}, but considering different impact parameters ($b$), that is varying the planet trajectory from an edge-on transit ($b=0$) to a grazing transit ($b=1$) in steps of 0.2, as outlined in the illustration in Fig. \ref{fig: Jup b}. This was done because previous studies have shown that the CLV curve is affected by planetary parameters, such as the planetary radius and impact parameter. The effect from the former is straightforward: the larger the radius, the stronger the CLV effect since the planet will block more light during the transit. On the other hand, the effect of different trajectories defined by the impact parameter is more complicated to evaluate and we will investigate it further in Sect. \ref{sec: diff b}.

The final goal of this simulation is to investigate how the CLV of the stellar lines influences the observed spectral line profile during the transit of a Jupiter-sized planet in front of a Sun-like star. This is done by evaluating how the relative flux  varies during the transit, that is at different orbital phases. Previous studies on the \ion{Na}{I} D lines (e.g., \citealt{Czesla2015}; \citealt{Yan2017}) investigated passbands of 0.75, 1.5 and $3.0\,\AA$, noting that the CLV effect is more pronounced in the narrower bandwidth, which is also where the planetary absorption is more prominent. 
We therefore selected the same three bandwidths to investigate how the line profile changes at different orbital phases during the transit, and with different impact parameters. All the out-of-transit relative flux is normalized to unity.

In the simulations, we considered 66 time steps between orbital phase $-0.0005$ and $+0.0005$, where the transit center occurs at phase zero. 
The CLV light curve, that is the relative flux as a function of the orbital phase, using the best-fit 3D non-LTE model for the \ion{Na}{I} D$_1$ line and an edge-on transit is shown in Fig. \ref{fig: CLV3DNLTE_diffbands}. In this figure, the relative flux in three bandwidths is highlighted in different colors. It can be noted that the CLV curves show a similar shape but the effect is more pronounced for the narrowest passband, that is $0.75\,\AA$ (green line in the figure), which is consistent with other studies. However, minimizing the CLV effect by choosing a larger passband might not be the best approach to use, since the absorption from the planet's atmosphere is usually concentrated in the core of the strong lines. Therefore, in order to better understand how the CLV varies among different trajectories, we analyzed only the flux in the $0.75\,\AA$ bandwidth, since the CLV effect is more prominent in this band. 

\subsubsection{Different impact parameters}\label{sec: diff b}
The CLV light curves for transits at different impact parameters are shown in Fig. \ref{fig: CLV3DNLTE_diffb} for the \ion{Na}{I} D$_1$, D$_2$ and \ion{K}{I} lines from top to bottom, respectively. 
In Fig. \ref{fig: CLV3DNLTE_diffb}, some interesting features in common between the three analyzed lines can be noticed. 
First of all, as expected, the CLV effect is stronger for an edge-on and almost edge-on transit (i.e., $b=0$ or $b=0.2$), represented by the blue and yellow solid lines. In particular, it is most pronounced around mid-transit, that is at phase 0, when the planet blocks the light from the very center of the stellar disk, where the intensity of the lines is stronger. 
Then, varying the impact parameter, the curve varies in a fairly uniform way until $b=0.6$ (red line), and the weakest curve occurs for a grazing transit at $b=1.0$ (brown line), as expected since the planet is obscuring only a small part of the stellar disk during the transit. However, in this specific case, an interesting effect occurs for the \ion{Na}{I} D$_1$ and D$_2$ lines, namely that the CLV curve resembles a planetary absorption profile. This means that for a Jupiter-sized planet orbiting a Sun-like star in this trajectory, the CLV effect can mimic a planetary atmospheric absorption, thus leading to a possible misattribution in the planetary elemental abundance of
Na if it is not properly corrected. This highlights the importance of properly modeling the stellar CLV in transmission spectroscopy studies.

It is worth noting that even if the CLV curves of \ion{Na}{I} D$_1$ and D$_2$ lines have similar shapes at different geometry, the values are still slightly different. For example, in the case of $b=0$ (blue line), the CLV curve of the \ion{Na}{I} D$_1$ line is about 0.1\% lower than the \ion{Na}{I} D$_2$ at mid-transit. And the difference between the curve of the \ion{Na}{I} D$_2$ and the \ion{K}{I} curve at the same point is even larger (about 0.27\%). Indeed, the two curves exhibit a completely opposite behavior. For $b=0$ to 0.6, the \ion{Na}{I} doublet shows a peak with positive values at mid-transit, whereas the peak for the \ion{K}{I} line occurs for negative values. This contrast arises from the distinct CLV of the lines (Fig. \ref{fig: EWvsmuNaI} and \ref{fig: EWvsmu7699}), with the \ion{Na}{I} D$_1$ line weakening as the $\mu$-angles decrease, while the \ion{K}{I} line strengthens toward the limb. 
The different CLV can be explained since the two lines have very different strengths (the \ion{Na}{I} D lines are on the damping part of the curve-of-growth whereas the \ion{K}{I} resonance line is on the flat, saturated part) and are forming at different layers in the solar atmosphere, characterized by different values of atmospheric depth-dependent quantities, such as temperature, density, and velocity fields. 
Therefore, since different spectral lines show different CLV light curves for the same geometry of the planet-star system, it is fundamental to model the CLV individually for each line of interest.

\subsubsection{Different stellar models}\label{sec: diff models}
The emergent intensity spectra adopted to simulate the stellar surface of the host star in \texttt{StarRotator} affect the resulting CLV curve. As also noticed in a recent work by \citet{Reiners2023}, the amplitude and shape of the curve indeed change with different models, since the center-to-limb line profiles are different (see Sect. \ref{sec: EW} and \ref{sec: shapes}). In order to evaluate the impact of using different stellar models, that is 1D instead of 3D or LTE instead of non-LTE, we performed the same simulations of Sect. \ref{sec: diff b} again but using as input the best-fit stellar spectra from \texttt{Balder} computed in 1D LTE, 1D non-LTE and 3D LTE (see Sect. \ref{sec: EW}). We also performed a simulation at $b=0$ using the line profiles of the SST data. 
In Fig. \ref{fig: CLV_diffmodels} we show the CLV light curves of the analyzed lines with different models and the SST data for an edge-on transit ($b=0$), computed in a band of $0.75\,\AA$, since it is the configuration with the strongest feature. The differences between the models depend on each individual line but there are several similar patterns.

%1D LTE vs 1D NLTE
\citet{Yan2017} noticed that, for the \ion{Na}{I} D lines, the CLV effect is less pronounced for a 1D non-LTE model compared to the corresponding 1D LTE model. In this work we find a similar result, that is clearly recognizable in the top panel of Fig. \ref{fig: CLV_diffmodels}, where the relative flux of the 1D LTE model (dashed red line) is larger by about 0.15\% than the corresponding 1D non-LTE model (dashed blue line). This is valid also for the \ion{K}{I} line in the bottom panel where the difference between the two 1D models is about 0.1\%.  

%1D NLTE vs 3D NLTE
For the models computed in non-LTE, in the \ion{Na}{I} D$_1$, the difference between a 3D (solid blue line) and a 1D (dashed blue line) atmosphere is that the former produces a less prominent CLV feature of about 0.07\%. 
On the other hand, the situation is reversed for the \ion{K}{I} line, where the 3D non-LTE model not only shows a more pronounced feature compared to the 1D non-LTE curve (of about 0.06\%) but also an inversion in the relative flux trend with respect to the 1D model, especially noticeable at mid-transit. This is due to the different CLV trend of the synthetic spectra, shown in Fig. \ref{fig: EWvsmu7699} in Sect. \ref{sec: EW}, where the 3D non-LTE model gets stronger at the limb, while all the other models exhibit line weakening for decreasing $\mu$-angles. Specifically, the weaker the line gets towards the limb, the larger the amplitude of the CLV curve will be at mid-transit. For both lines, the 1D LTE model is the one that decreases the most in equivalent width (Fig. \ref{fig: EWvsmuNaI} and \ref{fig: EWvsmu7699}), and therefore it is producing a higher peak in the CLV light curve in Fig. \ref{fig: CLV_diffmodels}. 

% 3D LTE vs 1D models
In all cases, the 3D LTE model (solid red line) lies in-between the 1D LTE and 1D non-LTE curves. The latter more closely resembles the 3D non-LTE model and the SST data. Consequently, from the analysis, it turns out that using a 1D non-LTE model is more accurate than using a 3D stellar atmosphere with an LTE line formation.

In Fig. \ref{fig: CLV_diffmodels} also the CLV light curve obtained using the SST line profiles is overplotted (open bullets). In both panels, it closely resembles the 3D non-LTE model, with a maximum difference at mid-transit of only 0.003\% for the \ion{Na}{I} D$_1$. The 3D non-LTE model of \ion{K}{I} performs almost as well, with a maximum difference of about 0.008\% at mid-transit, reflecting the slightly worse fit to the CLV curve (see Sect. \ref{sec: EW} and \ref{sec: shapes}). In any case, the agreement between the CLV curves of the SST data and the 3D non-LTE spectra is excellent, thus confirming that 3D non-LTE spectra are necessary to model the CLV effect and to get the most accurate results in transmission spectroscopy.

This analysis shows that different assumptions on the computation of the stellar synthetic spectra (1D or 3D, LTE or non-LTE) lead to different CLV curves and consequently to different estimates of the planetary elemental abundances. 
The differences in the CLV curves between models, even if small, are still important since they are of the same order of magnitude as a planetary absorption depth, which is very weak and typically between $0.01-0.5$\% (e.g., \citealt{Snellen2008}; \citealt{Yan2017}; \citealt{Mounzer2022}). Using a wrong CLV model could either mimic the presence of a species, thus leading to a false detection, or under(over)-estimate the abundance of a species in the planetary atmosphere. 
Therefore, a careful modeling as realistic and as precise as possible is required in order to correctly characterize exoplanet atmospheres.

\subsubsection{Rossiter McLaughlin effect}\label{sec: CLV+RM}
The Rossiter McLaughlin (RM) effect is an additional source of distortion of the stellar line profile that occurs during an exoplanetary transit and that could be as critical as the CLV effect, especially in fast-rotating stars like KELT-9 (\citealt{Hoeijmakers2019}). When an exoplanet transits across the stellar disk, it obscures different parts of the stellar surface, each having a local rotational velocity vector that could be pointing towards or away from the observer, depending on the transit phase. Consequently, red- or blue-shifted photons are removed from the total starlight, implying that the averaged wavelength is slightly red or blue-shifted as well. In order to investigate the combined effect of the CLV and the RM on the selected line profiles, we perform 

simulations of an edge-on transit ($b=0$) with a non-zero rotational velocity ($v_\mathrm{rot}$) using our 3D non-LTE synthetic spectra of the \ion{Na}{I} D$_1$ line. Specifically, one simulation assuming a slow-rotator like a Sun-like star, with a $v_\mathrm{rot}$ equal to the rotational velocity of the Sun, that is 2069.44 m s$^{-1}$ at the equator (\citealt{Meftah2014}), and another simulation for a fast-rotator (i.e., $P_\mathrm{rot} \leq 10$ days), assuming a $v_\mathrm{rot}= 10 \ \mathrm{km \ s^{-1}}$, which for a Sun-like star would correspond to a rotational period of about 5 days. In these simulations we assume that the star rotates as a solid body, meaning that differential rotation is not implemented in the code, and consequently, it is neglected in our analysis.

The results of these simulations are shown in Fig. \ref{fig: Dopplershadows} and \ref{fig: CLV+RM}, in comparison with the same planet-star configuration with the CLV effect only. The former figure shows the signal imprinted by the planet when transiting the stellar disk, also often referred to as "Doppler shadow". It usually represents the combination of the stellar CLV and RM, and this signal becomes stronger the faster the star rotates. In Fig. \ref{fig: CLV+RM}, the transmission curve computed in a bandwidth of $0.75\,\AA$ is shown for the three models with different stellar rotational velocities. As is apparent, the curve becomes asymmetric due to the RM effect. The asymmetry is more relevant for higher $v_\mathrm{rot}$ and the amplitude of the curve becomes larger as well. In the Sun case, the CLV is the dominant source of distortion of the stellar lines, and this is also valid for other slow-rotating Sun-like stars, such as HD 189733 (\citealt{Borsa2018}). On the other hand, in very fast-rotating stars, such as KELT-9 where the lines are very broadened due to the high rotational velocity ($v \sin(i) =111.4 \ \mathrm{km \ s^{-1}}$, \citealt{Gaudi2017}), the RM becomes significantly more important than the CLV.

\section{Discussion and Conclusions}\label{sec: conclusions}
In this work, we model the CLV of the \ion{Na}{I} D$_1$ $5896\,\AA$ line and \ion{K}{I} $7699\,\AA\,$ line using 3D hydrodynamic solar models and non-LTE line formation. We compare our synthetic spectra to solar observations from the high-resolution IAG Atlas as well as to SST/CRISP data taken in two different Campaigns in August 2022 and April 2023. We include in the comparison also 1D LTE, 1D non-LTE, and 3D LTE synthetic spectra. For the model atmosphere, we use ten snapshots from the \texttt{Stagger} solar simulation in 3D, and the \texttt{MARCS} solar model in 1D.
We investigate the line profile as well as the equivalent width ($W_\lambda$) of the lines as a function of the $\mu$-angle. The 3D non-LTE model represents the best match in both cases. It is able to reproduce the line shape from the center to the limb quite well and it is in reasonable agreement with the observed $W_\lambda$ within 6\% and 0.4\% at $\mu=0.1$ for the \ion{K}{I} and \ion{Na}{I} D$_1$ line, respectively. 1D models are symmetric by design therefore they are not able to reproduce the line asymmetries that are particularly strong in resonance lines. A 3D stellar atmosphere and a proper treatment of convection are required.
The LTE models typically produce lines with cores weaker than the observations but wings that are reasonably in agreement, at least for disk center line profiles. However, $W_\lambda$ is severely underestimated, with differences up to 37\% for the \ion{K}{I} line at the limb.

With the best-fit synthetic spectra, we also perform simulations of spectral lines of a Sun-like star transited by a Jupiter-sized planet without an atmosphere in order to investigate the impact of the CLV on the inference of planetary absorptions. In transmission spectroscopy the CLV effect can be of the same order of magnitude as the planetary absorption feature, therefore it is fundamental to adopt model spectra as accurate as possible. We compute transmission curves for different trajectories, or impact parameters ($b$), and evaluate the flux in three different bandwidths centered on the line cores. The LTE models, both with 1D and 3D atmospheres, produce a CLV curve with an amplitude larger than in non-LTE, thus overestimating the effect of CLV. In the presence of a planetary absorption, this would eventually result in an underestimate of the abundance of the analyzed species in the planetary atmosphere. In a recent work by \citet{Reiners2023}, they also found that the LTE curve amplitude is overpredicted when evaluating the CLV effect in a simulated planet-star system. They analyzed lines of several atomic species but they stress the fact that for \ion{Na}{I}, non-LTE effects are definitely non-negligible in order to reproduce observations. 
In this work, the differences between models, 1D vs 3D and LTE vs non-LTE, are relatively small but of the same order of magnitude as a planetary absorption depth. The CLV curves also greatly vary in shape for different geometries of the star-planet system and for different lines. Even the CLV curves of lines of the same species, that is \ion{Na}{I} D$_1$ and D$_2$ lines, are not identical. The CLV effect is indeed line-dependent and must be modeled line by line individually. We also perform a transit simulation with the SST line profiles and we find that the 3D non-LTE models are able to reproduce the CLV light curve of the observations extremely well.

Since also the planetary parameters, such as the planet radius and impact parameter, are crucial to correctly model the shape and strength of the CLV curve \citep{pietrow2023centertolimb}
, they need to be known with high accuracy. For instance, in our simulations, the CLV curve of the Na doublet resulting from a grazing transit ($b=1.0$) is very similar to a planetary absorption signature and therefore could be mistaken for a sodium detection if not properly corrected for. Nowadays, having accurate planetary parameters is possible thanks to the many ground-based as well as space-based telescopes performing high precision transit photometry, such as \textit{TESS} (\citealt{TESSRicker2014}), \textit{CHEOPS} (\citealt{CHEOPSref2021}) and also \textit{PLATO} (\citealt{PLATOref}) in the near future.

It is important to note that the simulation presented in this work represents a best-case scenario, given that only the quiet solar photosphere is considered. Inhomogeneities on the stellar surface due to active regions, such as starspots or plages, show a very different spectral line profile from the quiet regions (e.g., \citealt{Oranje1983}; \citealt{Dumusque2014}; \citealt{Pietrow2020};  \citealt{Rajhans2023}; \citealt{Pietrow2023c}; \citealt{Howard2023}; \citealt{Cretignier2023}). Observational data would allow for investigating the effect of magnetic fields on the line profiles. Hence, separate CLV should be measured in active regions in order to correctly incorporate them in the analysis of transmission spectra of active stars \citep{Chakraborty23}. Future work includes as well the modeling of synthetic spectra for active regions of different sizes and at various locations on the stellar disk.

%other stars
The differences in CLV transmission curves can be larger for different kinds of stars, especially for low-temperature stars where the effect is stronger as shown in \citet{Yan2017}. Consequently, a very good knowledge of the stellar parameters is as important as the precision of the planetary parameters in order to use a model stellar atmosphere as realistic as possible for the modeling of the CLV curves. A significant advance in this direction is given by the recent release of \textit{Gaia} DR3 (\citealt{Gaiaref2021}; \citealt{Katz2023}) and in the very near future by large spectroscopic surveys such as 4MOST (\citealt{deJong19}) and WEAVE (\citealt{Dalton18}; \citealt{Jin2023}), which will target millions of FGK stars providing accurate stellar parameters.

With the continuous improvement of the instrumentation used for transit spectroscopy, such as \textit{ESPRESSO} on VLT or \textit{HIRES} on ELT, the CLV effect will soon become crucial not only for the detection but especially for the characterization of exoplanetary atmospheres. At present, some facilities have already achieved the high signal-to-noise necessary for measuring the abundances of atmospheric species in Hot Jupiters and other gas giants. Several works have shown that using different models for the stellar spectrum can lead to very different results. For example, the case of the benchmark exoplanet HD 209458b in which the presence of Na was debated in the literature. Initial works detected it in transmission (e.g., \citealt{Charbonneau2002}; \citealt{Jensen2011}) but then later analysis argued that it was just an artifact due to the modeled stellar spectrum (\citealt{Casasayas2020, Casasayas2021}). This highlights the importance of using correct stellar models to evaluate the subtle effects like the CLV that could compromise the analysis of transit spectroscopy observations.

The Sun is the only star we can spatially resolve, therefore developing models able to reproduce solar observations is necessary to improve the CLV model of other stars as well, especially considering that transmission observations are expected to become more and more precise in the coming years. From our analysis, we can conclude that 1D models of Sun-like stars seem to overestimate the CLV signature in transmission curves. We can also say that, in accordance with previous works, the RM effect has an impact on the transmission curve as well which must be taken into account \citep{Reiners2023}. Specifically, it becomes stronger than the CLV effect for fast-rotating stars. For Sun-like stars instead, the CLV remains the main source of distortion of the stellar lines during transit.  

%%%----- Future perspectives -----%%%

However, it is difficult to make predictions about other kinds of stars or even about specific exoplanets, since every planet-star system is different and we show that different configurations lead to very different transmission curves. Consequently, future work includes the analysis of a well-known extensively-studied hot Jupiter, such as HD 189733b or HD 209458b, for which several high-resolution observations are already available. The goal is to apply a 3D non-LTE stellar model to evaluate the CLV effect of \ion{Na}{I} and \ion{K}{I} in these benchmark exoplanets and then compare the results with previous works in the literature (e.g., \citealt{Yan2017}; \citealt{Keles2019, Keles2020}; \citealt{Sicilia2022}). The methodology presented in this work can be extended as well to other FGK stars using 3D stellar atmospheres from the new full \texttt{Stagger} grid (Rodr\'iguez D\'iaz et al. submitted). It is worth noting that there are only a few studies where \ion{Na}{I} and \ion{K}{I} have been detected simultaneously (e.g., \citealt{Casasayas2021}). Given the opposite CLV curve of the resonance lines of the two ions, it would be interesting to observe those lines for more planet-star systems.

% other lines and elements:
The impact of a 3D non-LTE modeling of the CLV of the stellar lines needs to be studied in other species and other spectral lines as well. In particular for transmission spectroscopy, interesting species that have been detected in the atmosphere of several exoplanets include \ion{Ca}{I}, \ion{Ca}{II} (e.g., \citealt{Yan2019}), \ion{Fe}{I} (e.g., \citealt{Hoeijmakers2018}), \ion{Fe}{II} (e.g., \citealt{Bello2022}) and \ion{H}{I} Balmer lines (e.g., \citealt{Fossati2023}).

% CLV in low-res spectroscopy:
Finally, with the recent launch of the \textit{JWST} (\citealt{Barstow2015}), a new era for low-resolution transit spectroscopy has just begun. This space telescope will allow the atmospheric characterization of a wide range of exoplanets, from Hot Jupiters down to rocky planets and super-Earths. Low-resolution spectra are even more susceptible to the CLV effect in a wavelength-integrated sense than the high-resolution ones. Several studies show that high-resolution spectra, if adequately corrected for the RM and CLV effect, can be used to isolate the different spectral components and mitigate the stellar contamination in low-resolution spectra (e.g., \citealt{Genest2022}). 
 Therefore, exploring the effect of stellar CLV on low-resolution spectroscopy is an issue that has to be addressed in the near future, and this work is a step forward to achieve the best possible results in exoplanet science with \textit{JWST}.

%---------- acknowledgements ----------% 

\begin{acknowledgements} 
We thank the anonymous referee for their comments, which have improved the manuscript.
We thank A. Brandeker for the very useful comments on the first draft of the paper. We thank A. Reiners and M. Ellwarth for providing us with the \ion{Na}{I} D$_1$ and \ion{K}{I} line profiles before the publication of the IAG Atlas, and for the errors on $\mu$-angles. We also thank H. Reggiani for providing the potassium model atom that he developed.
GC and KL acknowledge funds from the Knut and Alice Wallenberg foundation. KL and CL acknowledge funds from the European Research Council (ERC) under the European Union’s Horizon 2020 research and innovation programme (Grant agreement No. 852977).
AMA acknowledges support from the Swedish Research Council (VR 2020-03940).
AP was supported at AIP by grants from the European Commission’s Horizon 2020 Program under grant agreements 824064 (ESCAPE -- European Science Cluster of Astronomy \& Particle Physics ESFRI Research Infrastructures) and 824135 (SOLARNET -- Integrating High Resolution Solar Physics). 
The computations were enabled by resources provided by the Swedish National Infrastructure for Computing (SNIC) at UPPMAX, partially funded by the Swedish Research Council through grant agreement no. 2018-05973. We thank the PDC Center for High Performance Computing, KTH Royal Institute of Technology, Sweden, for providing access to computational resources and support.
The Swedish 1-m Solar Telescope is operated on the island of La Palma by the Institute for Solar Physics of Stockholm University in the Spanish Observatorio del Roque de los Muchachos of the Instituto de Astrofísica de Canarias. The Institute for Solar Physics was supported by a grant for research infrastructures of national importance from the Swedish Research Council (registration number 2017-00625). The Swedish 1-m Solar Telescope, SST, is co-funded by the Swedish Research Council as a national research infrastructure (registration number 4.3-2021-00169).
This research has made use of NASA's Astrophysics Data System (ADS) bibliographic services. This work has made use of the VALD database, operated at Uppsala University, the Institute of Astronomy RAS in Moscow, and the University of Vienna. We acknowledge the community efforts devoted to the development of the following open-source packages that were used in this work: numpy (\url{numpy.org}), matplotlib (\url{matplotlib.org}) and astropy (\url{astropy.org}). We also used the CRISPEX analysis tool (\citealt{Vissers2012}) and the ISPy library (\citealt{ISPy2021}).
\end{acknowledgements}

%%%% BIBLIOGRAPHY %%%%
%\bibliographystyle{mnras}
%\bibliographystyle{aa}
\bibliographystyle{aa_url}
\bibliography{references} %because the bibliography file is called 'references.bib'

%---------- APPENDIX ----------% 

\appendix 
%%%% NEW PART %%%%
\section{Additional tables and figures}
%%%% Comparison between old and new Na atom %%%%%%%%
\begin{figure}[H]
    \centering
    \includegraphics[width=9cm]{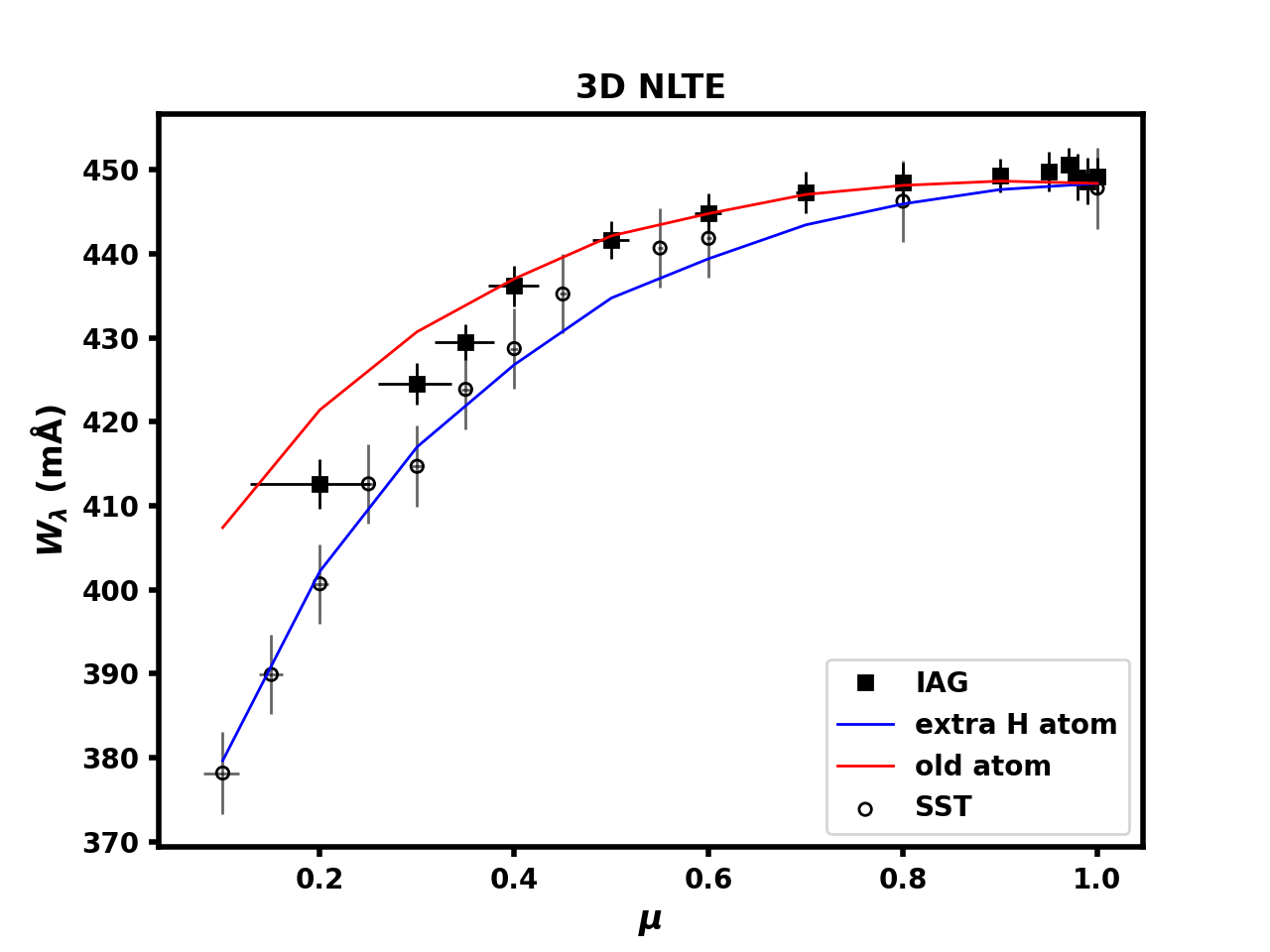}
    \caption{Comparison between 3D non-LTE models of the old Na atom and the new one with added H rate coefficients from \citet{Kaulakys1991}.} %from EW_Na_lines_extraH.ipynb
    \label{fig: atom-comparison}
\end{figure}
%%%%%%%%%%%%%%%%%%%%%%%%%%%%%%%%%%%%%%%%%%%%%%%%%%%%%%%
%%%%%%%%%%%%%%%%%%%%%%%%%%%%%%%%%%%%%%%%%%%%%%%%%%%%%%%%%%%%%%%%
\begin{table}[H] 
    \centering
    \caption{Equivalent widths (in m\AA) measured for the observed spectra of the SST data at different $\mu$-angles via direct integration of the line profiles. Note that the line profile of \ion{Na}{I} $5896\,\AA$ is not integrated completely but it is truncated on the right wing to exclude the blending due to a telluric line.} 
    \begin{tabular}{c c c} \hline \hline 
        $\mu$ &  $W_{5896}$ (m\AA) & $W_{7699}$ (m\AA)\\ \hline
         1.00 & $447.8 \pm 4.7$ & $157.5 \pm 3.3$\\
         0.80 & $446.3 \pm 4.7$ & $160.1 \pm 3.3$\\
         0.60 & $441.9 \pm 4.7$ & $166.2 \pm 3.4$\\
         0.55 & $440.7 \pm 4.7$ & $168.9 \pm 3.4$\\
         0.45 & $435.2 \pm 4.8$ & $171.3 \pm 3.4$\\
         0.40 & $428.7 \pm 4.8$ & $173.7 \pm 3.4$\\
         0.35 & $423.9 \pm 4.8$ & $175.6 \pm 3.4$\\
         0.30 & $414.7 \pm 4.7$ & $178.3 \pm 3.5$\\
         0.25 & $412.6 \pm 4.7$ & $181.2 \pm 3.4$\\
         0.20 & $400.7 \pm 4.7$ & $184.1 \pm 3.4$\\
         0.15 & $389.9 \pm 4.8$ & $188.2 \pm 3.4$\\
         0.10 & $378.2 \pm 4.9$ & $194.2 \pm 3.4 $ \\ \hline 
    \end{tabular}
    \label{tab: EWSST}
\end{table}
%%%%%%%%%%%%%%%%%%%%%%%%%%%%%%%%%%%%%%%%%%%%%%%%%
%%% Table with the observed EWs of the IAG Atlas %%%
\begin{table} %*
    \centering
    \caption{Equivalent widths (in m\AA) measured for the observed spectra of the IAG Atlas at different $\mu$-angles via direct integration over the wavelength range specified by $\lambda_\mathrm{int}$ in Table \ref{tab: atomicdata}. Note that the line profile of \ion{Na}{I} $5896\,\AA$ is not integrated completely but it is truncated on the right wing to exclude the blending due to a telluric line.}\label{tab: EWIAG}
    \small
    \begin{tabular}{c c c c c c} \hline \hline
     %& & & $W_\lambda$ (m\AA) &  & \\
    line & 5896 & 5688 & 6154 & 6160 & 7699\\ \hline 
    $\mu$ & & & $W_\lambda$ (m\AA) & & \\
    1.00 & $449.3 \pm 2.1$ & $122.1 \pm 1.3$ & $35.6 \pm 1.1$ & $55.7 \pm 1.2$ & $160.1 \pm 1.4$\\ 
    0.99 & $448.6 \pm 2.8$ & $122.1 \pm 1.7$ & $35.8 \pm 1.5$ & $55.8 \pm 1.5$ & $160.8 \pm 1.8$\\
    0.98 & $449.1 \pm 2.8$ & $122.5 \pm 1.7$ & $35.6 \pm 1.5$ & $55.7 \pm 1.5$ & $160.7 \pm 1.8$\\
    0.97 & $450.6 \pm 2.0$& $122.4 \pm 1.2$ & $35.7 \pm 1.1$ & $55.7 \pm 1.1$ & $160.4 \pm 1.3$\\
    0.95 & $449.8 \pm 2.3$& $122.2 \pm 1.4$ & $35.7 \pm 1.2$ & $55.5 \pm 1.3$ & $160.4 \pm 1.5$\\
    0.90 & $449.3 \pm 2.0$ & $121.9 \pm 1.2$ & $35.9 \pm 1.1$ & $55.6 \pm 1.1$ & $161.1 \pm 1.3$\\
    0.80 & $448.6 \pm 2.4$ & $121.2 \pm 1.5$ & $36.5 \pm 1.3$ & $55.9 \pm 1.3$ & $163.2 \pm 1.6$\\
    0.70 & $447.3 \pm 2.5$ & $120.7 \pm 1.6$ & $37.1 \pm 1.4$ & $56.4 \pm 1.4$ & $165.7 \pm 1.7$\\
    0.60 & $445.8 \pm 2.3$ & $120.3 \pm 1.5$ & $37.6 \pm 1.4$ & $56.8 \pm 1.4$ & $168.0 \pm 1.6$\\
    0.50 & $441.7 \pm 2.3$ & $120.1 \pm 1.5$ & $38.5 \pm 1.4$ & $57.4 \pm 1.4$ & $172.3 \pm 1.6$\\
    0.40 & $436.2 \pm 2.4$ & $119.7 \pm 1.6$ & $38.9 \pm 1.5$ & $57.7 \pm 1.5$ & $175.7 \pm 1.8$\\
    0.35 & $429.5 \pm 2.1$ & $119.6 \pm 1.4$ & $39.2 \pm 1.3$ & $57.8 \pm 1.3$ & $178.0 \pm 1.6$\\
    0.30 & $424.5 \pm 2.5$ & $119.3 \pm 1.7$ & $39.5 \pm 1.6$ & $58.1 \pm 1.6$ & $181.0 \pm 1.9$\\
    0.20 & $412.6 \pm 2.9$ & $119.1 \pm 2.0$ & $40.1 \pm 1.9$ & $58.5 \pm 1.9$ & $186.3 \pm 2.4$\\ \hline
    \end{tabular}
\end{table}
%%%%%%%%%%%%%%%%%%%%%%%%%%%%%%%%%%%%%%%%%%%%%%%%%%%%%
\begin{figure*}
\resizebox{\hsize}{!}
          {\includegraphics[width=5cm]{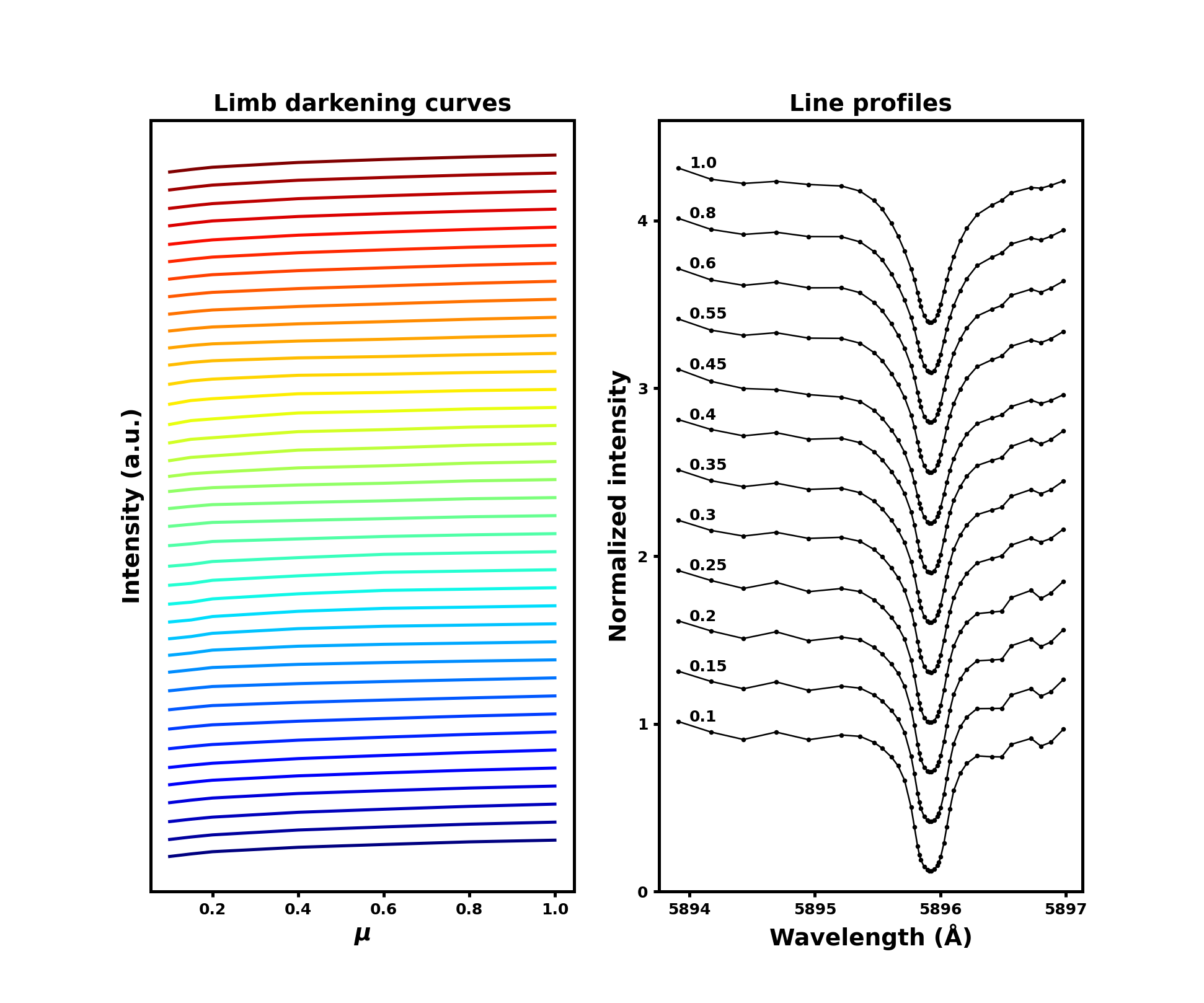}}
          \caption{\ion{Na}{I} D$_1$ data taken in Campaign I. \textbf{Left:} The limb darkening data of each wavelength point in the \ion{Na}{I} D$_1$ line on an arbitrary intensity scale. The curves are ordered by wavelength, with the top (red) curve representing the red continuum, the middle (green) curve representing the line core, and the bottom (blue) curve representing the blue continuum. \textbf{Right:} Normalized intensity of the averaged line profiles at different $\mu$-angles. Spectra for $\mu \geq 0.15$ have been incrementally offset vertically by $+0.3$ for clarity.}
          \label{fig: profilesNaD1}
\end{figure*}
%%%%%%%%%%%%%%%%%%%%%%%%%%%%%%%%%%%%%%%%%%%%%%%%%%%
%%%% Line profiles and limb darkening curves of K I SST data %%%%
% K I 
\begin{figure*}
\resizebox{\hsize}{!}
          {\includegraphics[width=5cm]{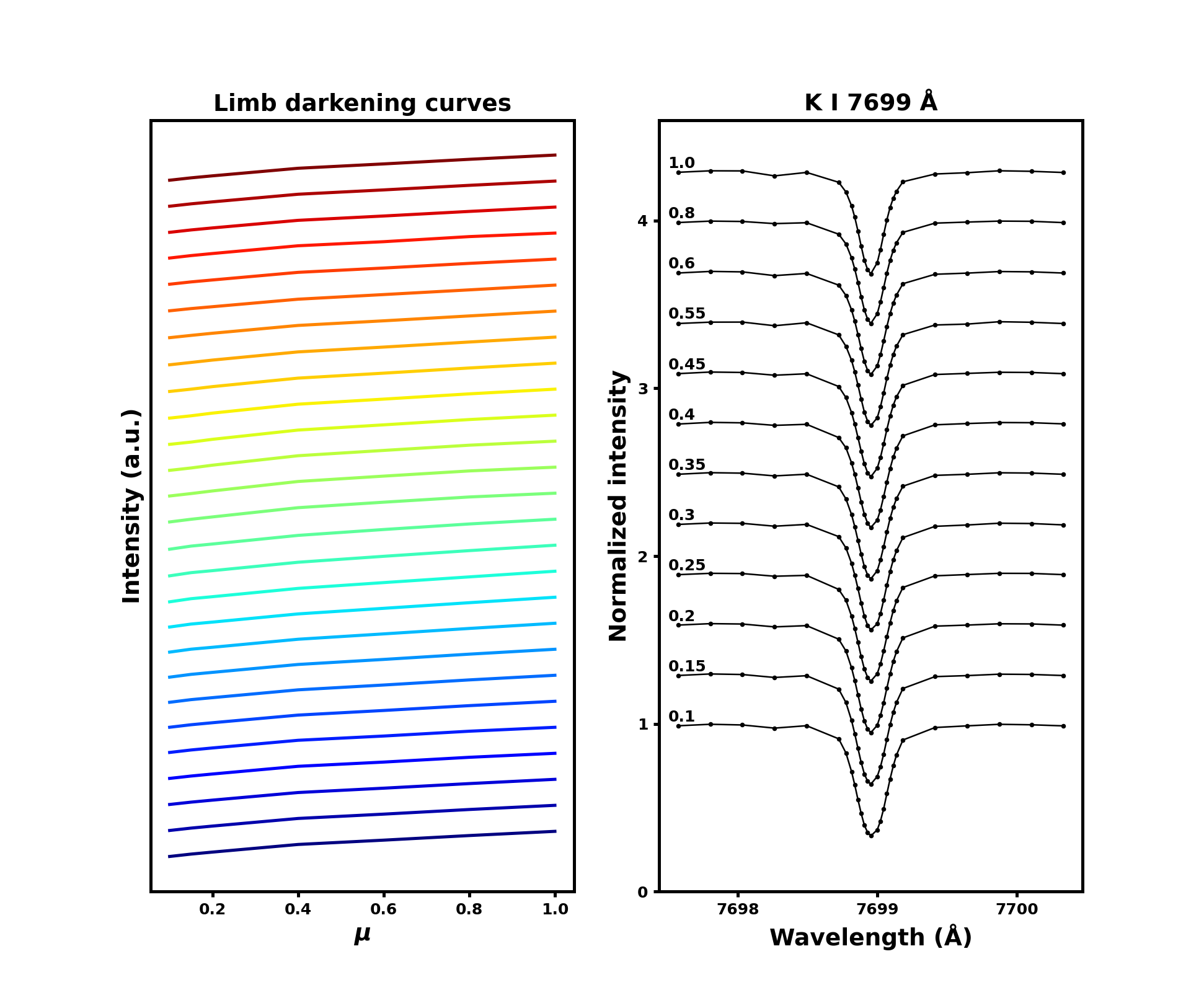}}
          \caption{\ion{K}{I} data taken in Campaign II. \textbf{Left:} The limb darkening data of each wavelength point in the \ion{K}{I} $7699\,\AA$ line on an arbitrary intensity scale. The curves are ordered by wavelength, with the top (red) curve representing the red continuum, the middle (green) curve representing the line core, and the bottom (blue) curve representing the blue continuum. \textbf{Right:} Normalized intensity of the averaged line profiles at different $\mu$-angles. Spectra for $\mu \geq 0.15$ have been incrementally offset vertically by $+0.3$ for clarity.}
          \label{fig: profilesK}
\end{figure*}

\begin{figure*}
 \centering
   \resizebox{\hsize}{!}{\includegraphics[width=9cm]{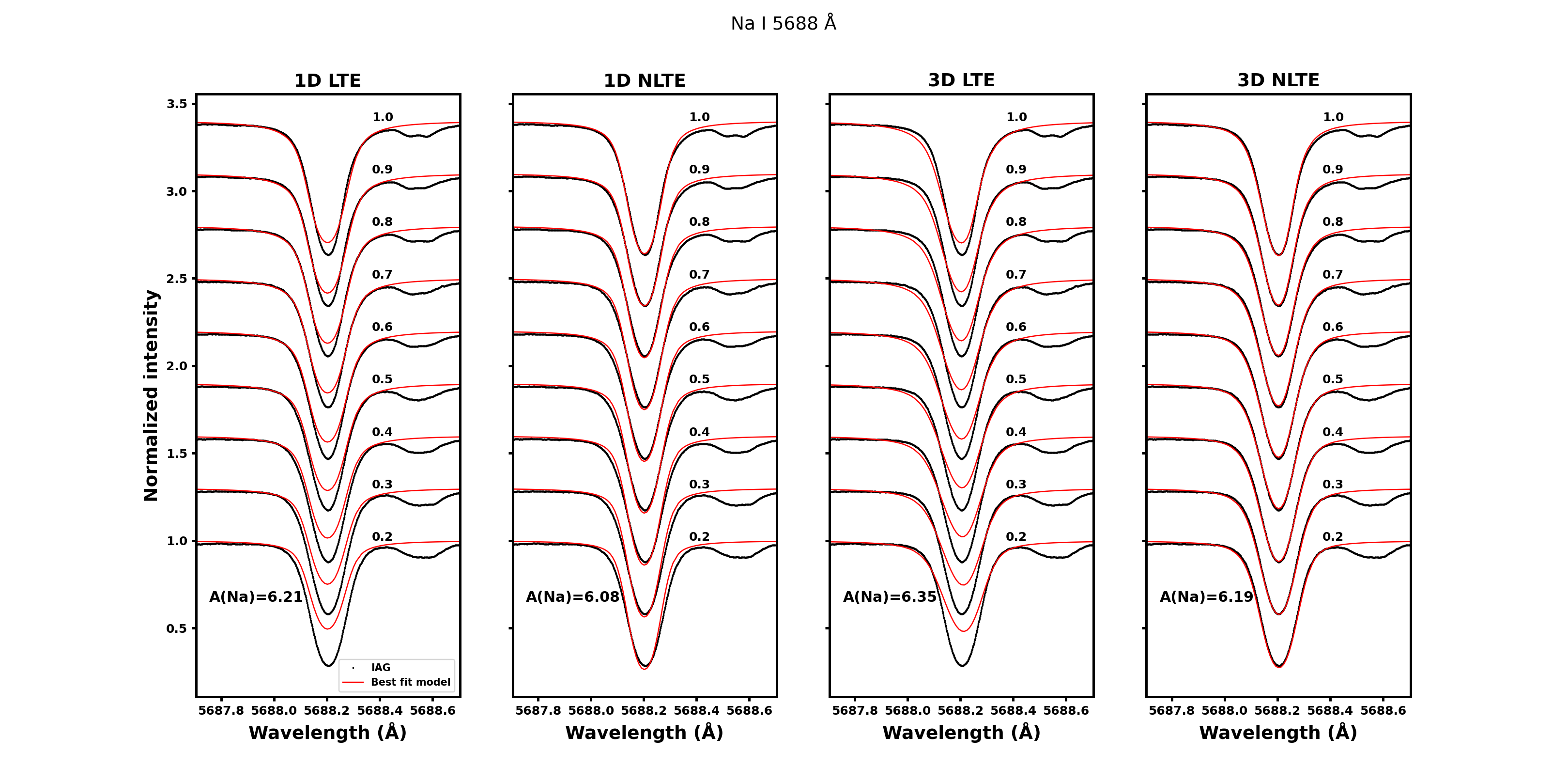}}
      \caption{Normalized observed (black dots) intensity and synthetic (red lines) center-to-limb profiles for the \ion{Na}{I} $5688\,\AA$ line, where the numbers above each spectrum correspond to the $\mu$-angle. In the left corner, the calibrated abundance is shown. Spectra for $\mu \geq 0.3 $ have been incrementally offset vertically by $+0.3$ for clarity.}
         \label{fig: line5688}
   \end{figure*}
\begin{figure*}
 \centering
   \resizebox{\hsize}{!}{\includegraphics[width=9cm]{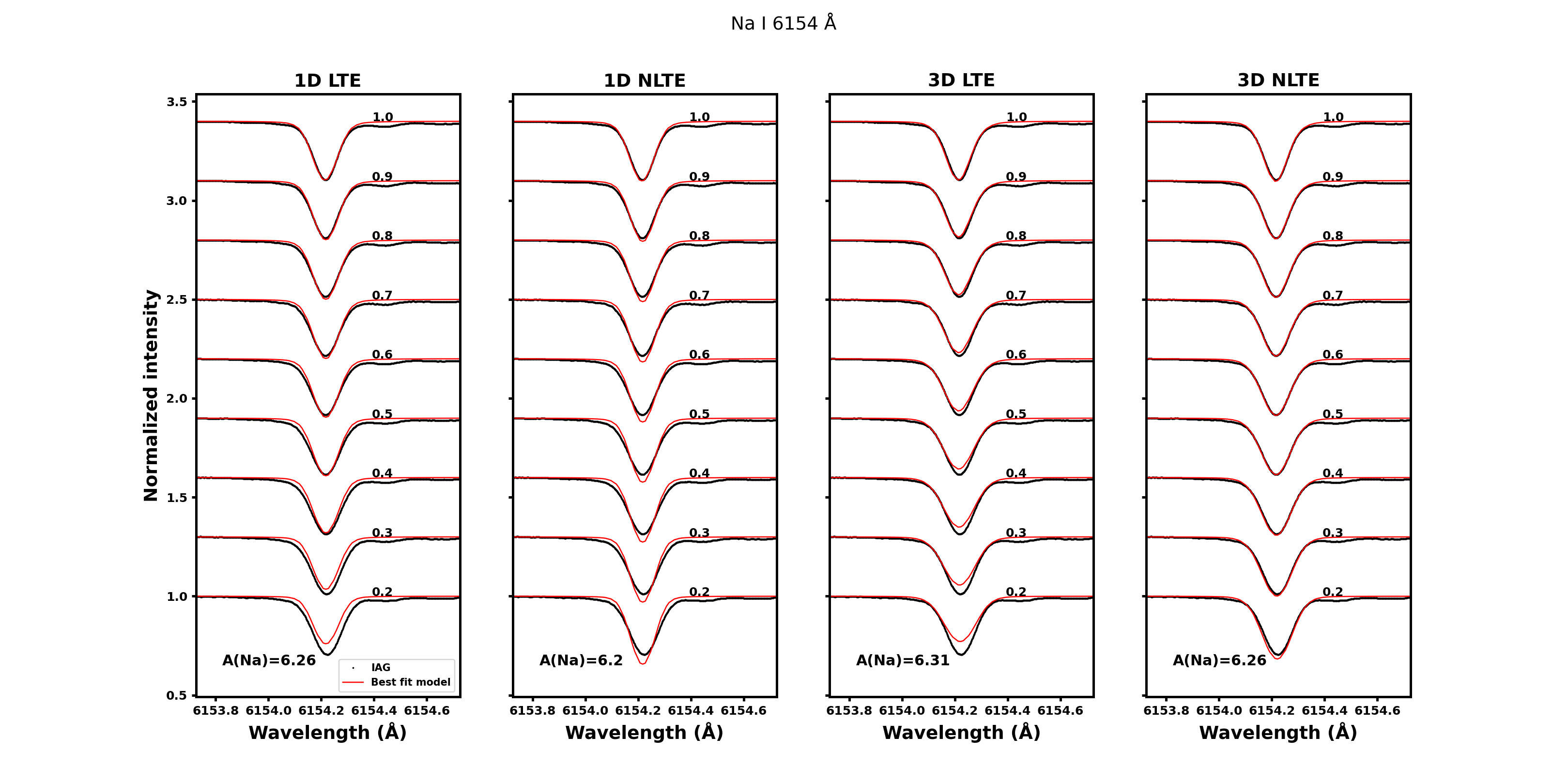}}
      \caption{Normalized observed (black dots) intensity and synthetic (red lines) center-to-limb profiles for the \ion{Na}{I} $6154\,\AA$ line, where the numbers above each spectrum correspond to the $\mu$-angle. In the left corner, the calibrated abundance is shown. Spectra for $\mu \geq 0.3 $ have been incrementally offset vertically by $+0.3$ for clarity.}
         \label{fig: line6154}
   \end{figure*}
\begin{figure*}
 \centering
   \resizebox{\hsize}{!}{\includegraphics[width=9cm]{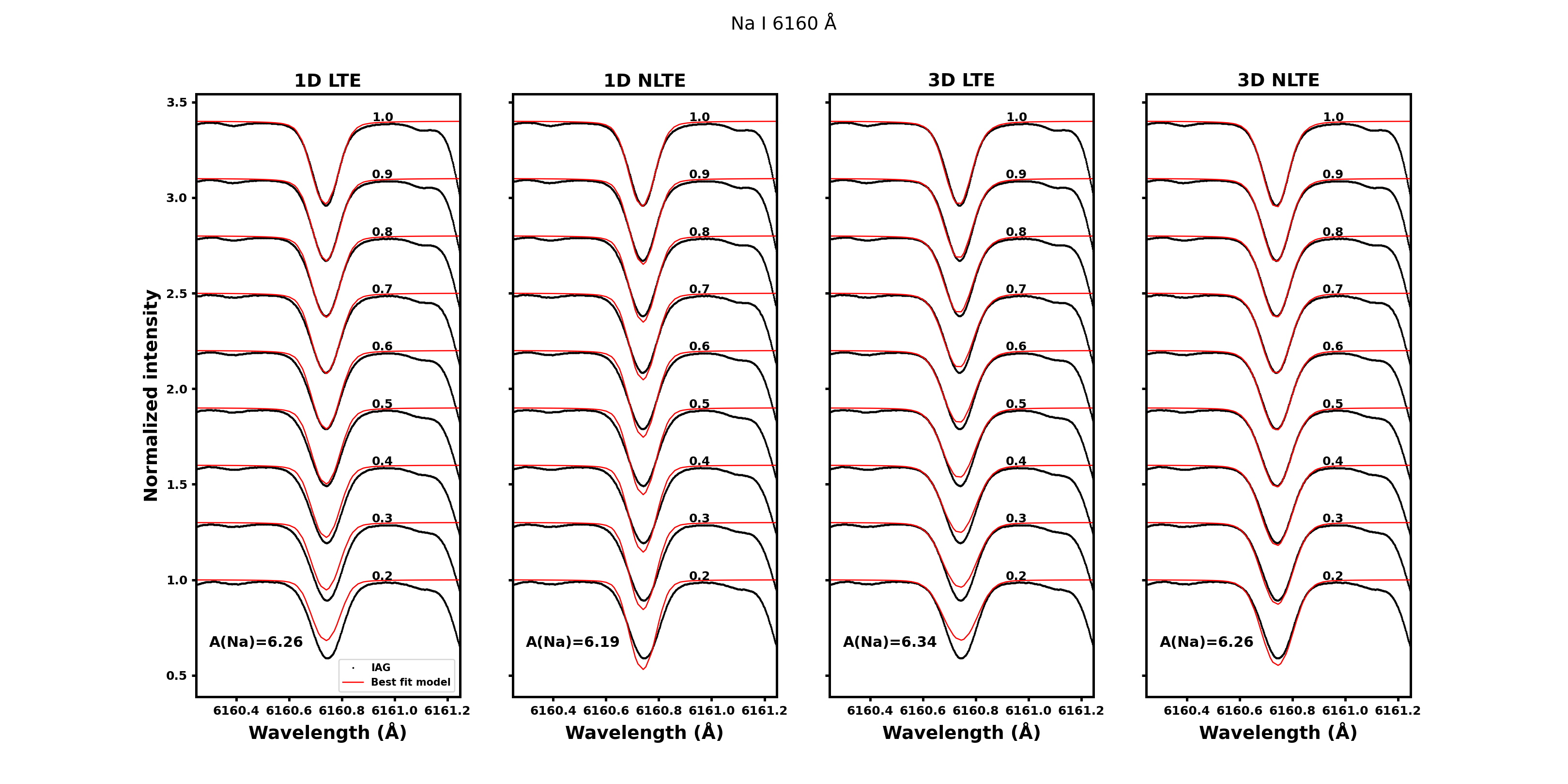}}
      \caption{Normalized observed (black dots) intensity and synthetic (red lines) center-to-limb profiles for the \ion{Na}{I} $6160\,\AA$ line, where the numbers above each spectrum correspond to the $\mu$-angle. In the left corner, the calibrated abundance is shown. Spectra for $\mu \geq 0.3 $ have been incrementally offset vertically by $+0.3$ for clarity.}
         \label{fig: line6160}
   \end{figure*}
%%%%% Na D1 and K I at IAG original resolution %%%%%%%%%%
\begin{figure*}
 \centering
   \resizebox{\hsize}{!}{\includegraphics[width=9cm]{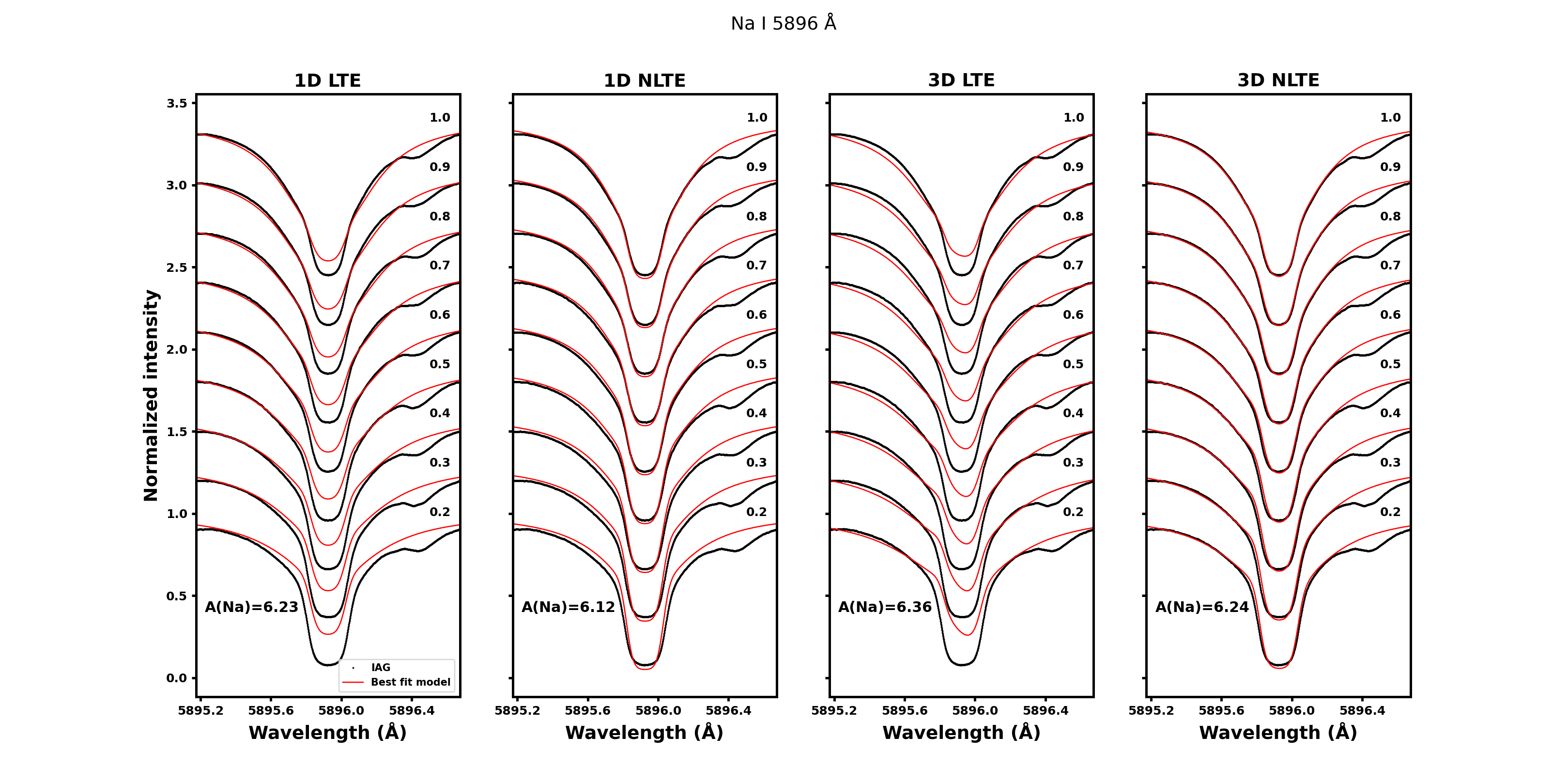}} 
      \caption{Normalized observed (black dots) intensity from the IAG Atlas and synthetic (red lines) center-to-limb profiles for the \ion{Na}{I} D$_1$ at $5896\,\AA$ line, where the numbers above each spectrum correspond to the $\mu$-angle. In the left corner, the calibrated abundance is shown. Spectra for $\mu \geq 0.3 $ have been incrementally offset vertically by $+0.3$ for clarity.}
         \label{fig: line5896IAG}
   \end{figure*}
   \begin{figure*}
 \centering
   \resizebox{\hsize}{!}{\includegraphics[width=9cm]{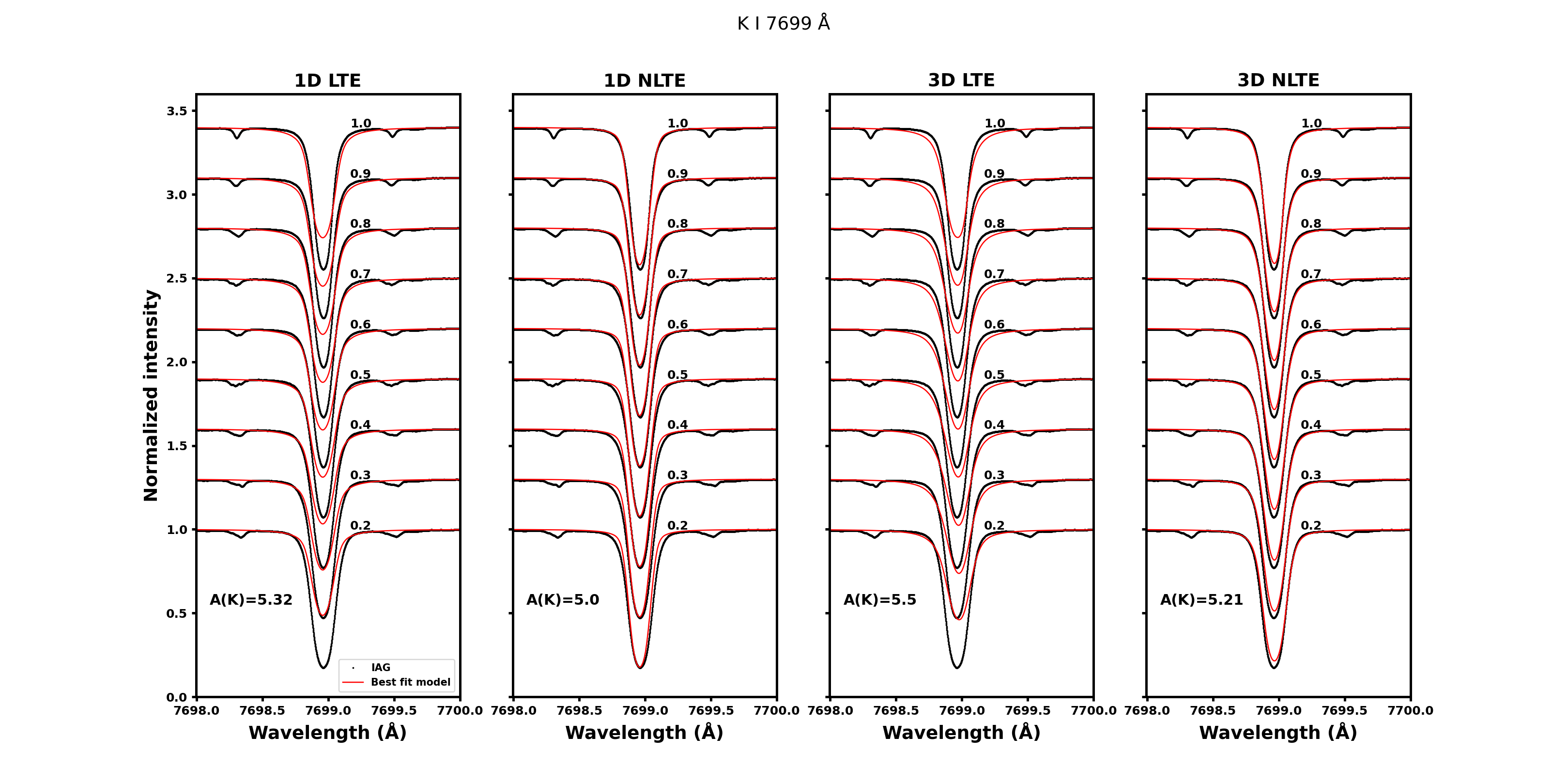}}
      \caption{Normalized observed (black dots) intensity from the IAG Atlas and synthetic (red lines) center-to-limb profiles for the \ion{K}{I} $7699\,\AA$ line, where the numbers above each spectrum correspond to the $\mu$-angle. In the left corner, the calibrated abundance is shown. Spectra for $\mu \geq 0.3 $ have been incrementally offset vertically by $+0.3$ for clarity.}
         \label{fig: line7699IAG}
   \end{figure*}
%%%%%%%%%%%%%%%%%%%%%%%%%%%%%%%%%%%%%%%%%%%%%%%%%%%%%%%%%%%%%%%%%%%%%%


%\begin{thebibliography}{}

%  \bibitem[Baker(1966)]{baker} Baker, N. 1966,in Stellar Evolution,ed.\ R. F. Stein,\& A. G. W. Cameron Plenum, New York) 333

%   \bibitem[Balluch(1988)]{balluch} Balluch, M. 1988,  A\&A, 200, 58

%\end{thebibliography}

\end{document}